\documentclass[twocolumn]{aastex631}

\begin{document}

\title{A Redshift Survey of the Coma Cluster (A1656):
Understanding the Nature of Subhalos in the Weak-lensing Map}

\author{Wooseok Kang}
\affiliation{Astronomy Program, Department of Physics and Astronomy, Seoul National University, 1 Gwanak-ro, Gwanak-gu, Seoul 08826, Republic of Korea}
\affiliation{Department of Physics, Korea Advanced Institute of Science and Technology (KAIST), 291 Daehak-ro, Yuseong-gu, Daejeon 34141, Republic of Korea}

\author{Ho Seong Hwang}
\affiliation{Astronomy Program, Department of Physics and Astronomy, Seoul National University, 1 Gwanak-ro, Gwanak-gu, Seoul 08826, Republic of Korea}
\affiliation{SNU Astronomy Research Center, Seoul National University, 1 Gwanak-ro, Gwanak-gu, Seoul 08826, Republic of Korea}
\affiliation{Australian Astronomical Optics - Macquarie University, 105 Delhi Road, North Ryde, NSW 2113, Australia}

\author{Nobuhiro Okabe}
\affiliation{Physics Program, Graduate School of Advanced Science and Engineering, Hiroshima University, 1-3-1 Kagamiyama, Higashi-Hiroshima, Hiroshima 739-8526, Japan}
\affiliation{Hiroshima Astrophysical Science Center, Hiroshima University, 1-3-1 Kagamiyama, Higashi-Hiroshima, Hiroshima 739-8526, Japan}
\affiliation{Core Research for Energetic Universe, Hiroshima University, 1-3-1, Kagamiyama, Higashi-Hiroshima, Hiroshima 739-8526, Japan}

\author{Changbom Park}
\affiliation{Korea Institute for Advanced Study, 85 Hoegi-ro, Dongdaemun-gu, Seoul 02455, Republic of Korea}

\begin{abstract}

    We study the physical properties of weak-lensing subhalos in the Coma cluster of galaxies using data from galaxy redshift surveys.
    The data include \textcolor{black}{12989} galaxies with measured spectroscopic redshifts (\textcolor{black}{2184} from our MMT/Hectospec observation and \textcolor{black}{10807} from the literature).
    The $r$-band magnitude limit at which the differential spectroscopic completeness drops below 50\% is \textcolor{black}{20.2} mag, which is spatially uniform in a region of 4.5 deg$^{2}$ where the weak-lensing map of Okabe et al. (2014) exists.
    We identify \textcolor{black}{1337} member galaxies in this field and use them to understand the nature of 32 subhalos detected in the weak-lensing analysis.
    We use Gaussian Mixture Modeling (GMM) in the line-of-sight velocity domain to measure the mean velocity, the velocity dispersion, and the number of subhalo galaxies by mitigating the contamination from the interloping galaxies.
    Using subhalo properties calculated with GMM, we find no significant difference in the redshift space distribution between the cluster member galaxies and subhalos.
    We find that the weak-lensing mass shows strong correlations with the number of subhalo member galaxies, velocity dispersion, and dynamical mass of subhalos with power-law slopes of $0.54^{+0.16}_{-0.15}$, $0.93^{+0.35}_{-0.32}$, and $0.50^{+0.31}_{-0.18}$, respectively.
    The slope of the mass--velocity dispersion relation of the weak-lensing subhalos appears shallower than that of the galaxy clusters, galaxy groups, and individual galaxies.
    These results suggest that the combination of redshift surveys with weak-lensing maps can be a powerful tool for better understanding the nature of subhalos in clusters.
    
\end{abstract}


\section{Introduction} \label{sec:intro}
    According to the current standard cosmological model (i.e., $\Lambda$ cold dark matter; $\Lambda$CDM), dark matter collapses from the initial density fluctuation to form dark matter halos \citep{Benson2010}; these become the nests of visible galaxies that are formed from the gas falling into the halos.
    Understanding the connection between the dark matter halos (or subhalos within halos; e.g., \citealt{Kim2008}) and galaxies is one of the key issues in cosmology and structure formation \citep{Hong2016, Wechsler2018}.
      
    In particular, the abundance of halos/subhalos (e.g., the number of halos/subhalos per unit mass per unit volume at redshift $z$, $\mathrm{d} n(M,z)/\mathrm{d}M$) has been a powerful tool for constraining the cosmological parameters \citep{Allen2011}.
    This halo/subhalo mass function can be modeled analytically (e.g., \citealt{Press1974, Zentner2007}) and numerically (e.g., \citealt{DeLucia2004, Kim2006, Springel2008, Klypin2011}).
    In the $\Lambda$CDM scheme, it is predicted to follow a power law of $\mathrm{d}n(M,z)/\mathrm{d}M \propto M^{-\alpha}$ with $\alpha\approx-1.9$ (see \citealt{Bullock2017} for more details).
    Comparing this with the observed mass function can validate the dark matter models including warm or self-interacting ones (e.g., \citealt{Vogelsberger2012, Lovell2020}).
    
    For example, \citet{Schwinn2017} suggested that the extraordinary amount of substructure in the field of Abell 2744 identified in \citet{Jauzac2016} appears inconsistent with the abundance and distribution of cluster subhalos in the Millennium XXL simulation \citep{Angulo2012}.
    However, this apparent discrepancy between observations and simulations can disappear once they use high-resolution simulations with careful calculation of subhalo masses \citep{Mao2018}: i.e., no tension with the $\Lambda$CDM cosmology.

    In addition to testing the cosmological model, the detection and mass measurement of subhalos can also provide information about the mass growth history of individual clusters.
    According to the current hierarchical merger scenario, small-scale structures form first and then merge to create larger structures, such as galaxy clusters.
    Thus, subhalos within clusters can serve as direct evidence of a past merger event.
    For example, \citet{HyeongHan2024_Perseus} found a subhalo within the Perseus cluster and measured its mass using weak-lensing analysis.
    They proposed an off-axis major merger scenario based on this subhalo and the distribution of the intracluster medium.
    The NGC 4839 group within the Coma cluster is also a well-known infalling system.
    The infall scenario of the NGC 4839 group has been actively studied using X-ray observations (e.g., \citealp{Neumann2001AA, Lyskova2019, Mirakhor2023}).
    
    Despite the importance of subhalos, it has been difficult to observationally detect subhalos without relying on visible galaxies.
    Weak-lensing analysis of galaxy clusters is useful for identifying not only large-scale structures around clusters (e.g., \citealt{HyeongHan2024_Coma}), but also small-scale subhalos in clusters (e.g., \citealt{Okabe2014}).
    However, it is still not easy to resolve small subhalos, which are crucial for determining the power-law slope of the subhalo mass function.
    In addition, the weak-lensing signal accounts for the mass projected along the line of sight \citep{Hoekstra2001}, and there could be non-negligible contamination in the identification of subhalos associated with galaxy clusters.
    We therefore conduct an extensive redshift survey for the Coma cluster to directly examine the nature of subhalos identified by the weak-lensing analysis of \citet{Okabe2014}.
    The comparison of the structures identified from weak lensing and redshift surveys has successfully demonstrated the importance of redshift surveys for better understanding the spatial distribution of dark matter in the fields \citep{Geller2005, Geller2010} and in cluster regions (\citealt{Geller2014_A383, Hwang2014, Liu2018}; see also \citealt{Shin2022} for the analysis of numerical simulations).
    This comparison can also result in an interesting discussion on dark substructures (e.g., \citealp{Clowe2012, Jee2014}) and dark galaxies (\citealp{Lee2024, kwon25alfalfa}).
    There have been some attempts to understand the nature of subhalos in the Coma cluster, but mainly with X-ray observations (e.g. \citealt{AndradeSantos2013, Sasaki2015, Sasaki2016}).
    This study would be the first of its kind to systematically investigate the nature of subhalos in the Coma cluster by combining weak-lensing analysis and redshift surveys.
     
    The paper is structured as follows.
    In Section \ref{sec:data}, we describe the data we use for the analysis of the Coma cluster.
    We present our results in Section \ref{sec:result}.
    In Sections \ref{sec:discuss} and \ref{sec:conclusion}, we discuss and conclude our study, respectively.
    Throughout the paper, we assume a $\Lambda$CDM cosmology with $H_{0}=100h \ \mathrm{km \ s^{-1} \ Mpc^{-1}}$, $\Omega_{\Lambda}=0.7$, and $\Omega_{m}=0.3$.

\section{Data}  \label{sec:data}
    \subsection{Photometric Data}   \label{sec:data:phot}
    We use the Sloan Digital Sky Survey (SDSS) Data Release 17 (DR17; \citealt{Abdurrouf2022_SDSSDR17}) photometric catalog as the basis for compiling our redshift catalog.
    The SDSS photometric catalog contains the right ascension, declination, apparent magnitude in each band, and a flag indicating whether an object is likely to be a point source, among other useful data.
    For spectroscopic observations, we prioritized brighter galaxies\footnote{We mainly observed extended sources as provided by the SDSS database (i.e., photometric flag of \texttt{p\_probpsf} equal to zero). However, we manually included 12 point sources in the observations to search for compact galaxies in the Coma cluster (to be discussed in the next section).} in the $r$-band over fainter galaxies for target selection without imposing other criteria, such as color.
    This strategy has been proven to be useful in previous studies (e.g., \citealp{Geller2014_A383}) to obtain unbiased samples of cluster galaxies.
    
    \subsection{Spectroscopic Data} \label{sec:data:spec}
    
        \begin{deluxetable*}{ccccccc}[!htb]
            \label{tab:obs_field}
            \caption{Summary of MMT/Hectospec observation fields.}
            \tablehead{
                \colhead{Field ID} & \colhead{R.A.}      & \colhead{Decl.}     & \colhead{Date} & \colhead{Exposure} & \colhead{Number of Targets} & \colhead{Number of Redshifts} \\ 
                \colhead{}         & \colhead{(\arcdeg)} & \colhead{(\arcdeg)} & \colhead{}     & \colhead{(min)}    & \colhead{} & \colhead{}
            }
            \startdata
                COMAa\_1    & 194.964425 & 27.979032 & 2014 Apr  2 & 45.0 & 261 & 233 \\
                COMAb15\_1  & 194.282042 & 27.481113 & 2015 May 27 & 45.0 & 255 & 117 \\
                COMAb15\_2  & 194.265246 & 28.010580 & 2015 May 27 & 45.0 & 260 & 141 \\
                COMAb15\_3  & 195.022046 & 27.509447 & 2015 Jun  2 & 60.0 & 258 & 51  \\
                COMAb15\_4  & 195.021417 & 27.999956 & 2015 Jun  3 & 60.0 & 259 & 73  \\
                COMAa17\_1  & 195.238037 & 27.386894 & 2017 Feb 24 & 75.0 & 258 & 253 \\
                COMAa17\_2  & 194.091208 & 27.375820 & 2017 Apr 19 & 60.0 & 258 & 240 \\
                COMAa17\_3  & 194.039125 & 28.190336 & 2017 Apr 26 & 60.0 & 247 & 228 \\
                COMAa17\_4  & 195.265417 & 28.153091 & 2017 Apr 30 & 60.0 & 248 & 223 \\
                COMAa17\_5  & 194.011787 & 28.094540 & 2017 May  1 & 60.0 & 243 & 221 \\
                COMAb18\_1  & 195.167254 & 27.251417 & 2018 May 20 & 54.0 & 249 & 222 \\
                COMAa19\_1  & 194.141337 & 27.250156 & 2019 Apr 28 & 60.0 & 252 & 235 \\
            \enddata
        \end{deluxetable*}

        \begin{figure*}[!ht]
            \centering
            \includegraphics[width=0.62\linewidth]{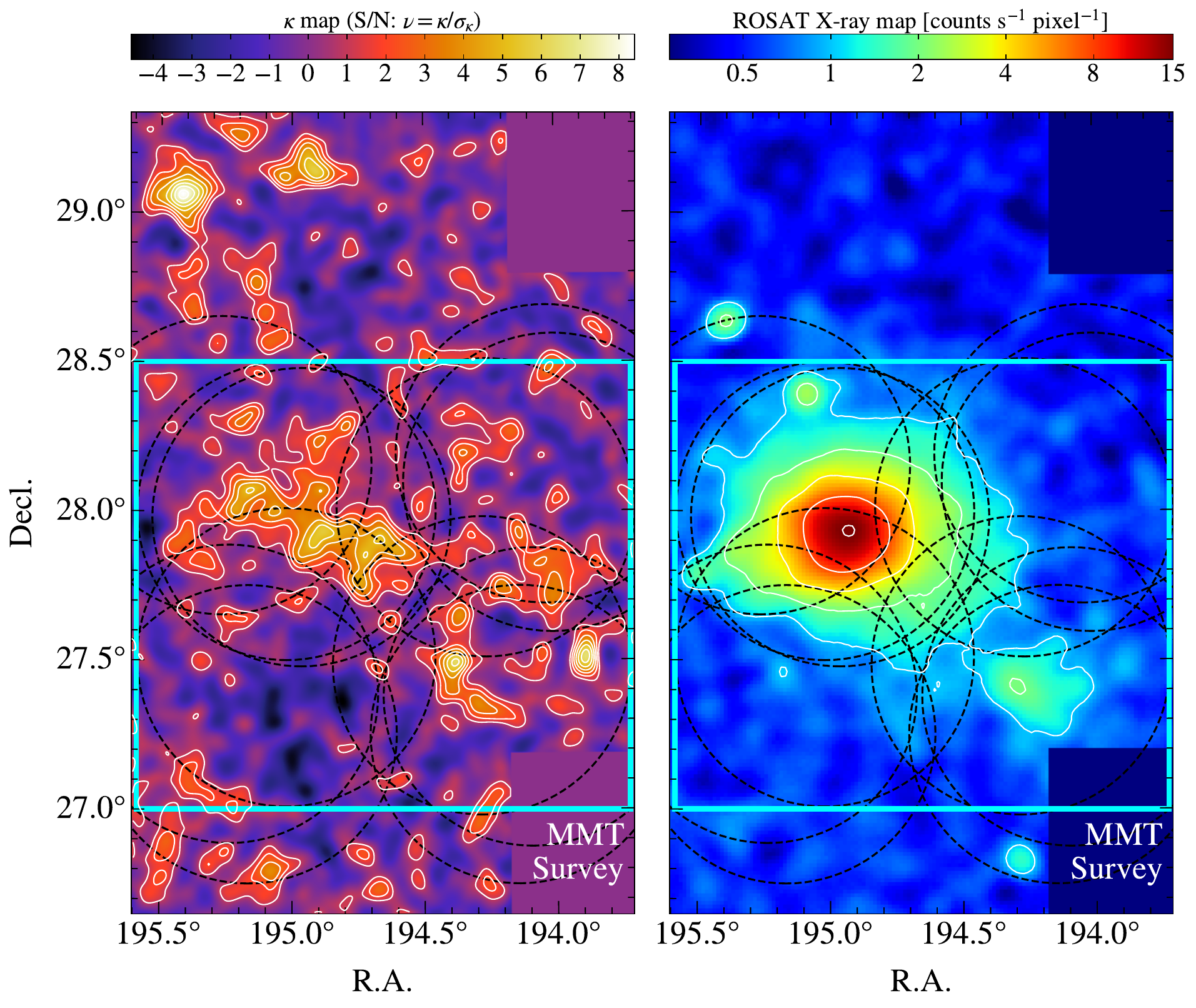}
            \caption{MMT/Hectospec survey region (cyan box) with the fields of view (black dashed circles).
            Left: Weak-lensing $\kappa$ map, in units of signal-to-noise adopted from \citet{Okabe2014}. 
            Right: X-ray surface brightness observed by \textit{ROSAT}.
            }
            \label{fig:fov}
        \end{figure*}

        \begin{deluxetable*}{ccccccccc}[!htb]
            \label{tab:catalog}
                \caption{Redshift Catalog of Objects within $132\arcmin$ of the Coma Cluster Center}
                \tablehead{
                    \colhead{ID} & \colhead{SDSS ObjID} & \colhead{R.A.} & \colhead{Decl.} & \colhead{$m_{r,\rm{Petro,0}}$} & \colhead{Point Source\tablenotemark{a}} & \colhead{$z$} & \colhead{$z$ Source\tablenotemark{b}} & \colhead{Member\tablenotemark{c}} \\
                    \colhead{} & \colhead{} & \colhead{(\arcdeg)} & \colhead{(\arcdeg)} & \colhead{(mag)} & \colhead{} & \colhead{} & \colhead{} & \colhead{}
                }

                \startdata
                    1	& 1237667444047741744 &	192.506759 &	28.021604 &	22.858202 &	0 &	 1.25812 $\pm$ 0.00005 &	3 &	0 \\
                    2	& 1237667444047741737 &	192.508991 &	28.124564 &	20.654058 &	0 &	 0.53041 $\pm$ 0.00014 &	2 &	0 \\
                    3	& 1237667323796783290 &	192.509175 &	27.888838 &	17.812452 &	1 &	-0.00004 $\pm$ 0.00001 &	3 &	0 \\
                    4	& 1237667444047741046 &	192.509191 &	27.932472 &	18.049040 &	1 &	-0.00014 $\pm$ 0.00001 &	3 &	0 \\
                    5	& 1237667324333654391 &	192.512303 &	28.151719 &	20.713013 &	0 &	 0.26260 $\pm$ 0.00001 &	3 &	0 \\
                    6	& 1237667323796783437 &	192.513204 &	27.787350 &	21.938515 &	1 &	 2.57419 $\pm$ 0.00034 &	2 &	0 \\
                    7	& 1237667323796783435 &	192.516752 &	27.863187 &	20.901819 &	1 &	 0.99865 $\pm$ 0.00012 &	3 &	0 \\
                    8	& 1237667324333654076 &	192.517496 &	28.252774 &	18.796087 &	1 &	 0.00005 $\pm$ 0.00001 &	3 &	0 \\
                    9	& 1237667324333654106 &	192.518737 &	28.224767 &	19.539011 &	1 &	 0.00029 $\pm$ 0.00001 &	3 &	0 \\
                    10  & 1237667324333654387 &	192.519747 &	28.268707 &	19.916807 &	0 &	 0.19558 $\pm$ 0.00006 &	3 &	0 \\
                \enddata
                \tablecomments{This table is available in its entirety in a machine-readable form.}
                \tablenotetext{a}{(0) Extended source, (1) Point source.}
                \tablenotetext{b}{(1) This work,
                                  (2) SDSS,
                                  (3) \citet[DESI EDR]{DESIEDR2024},
                                  (4) \citet{Koo1986},
                                  (5) \citet{Crampton1987},
                                  (6) \citet{Hewitt1989},
                                  (7) \citet{Boroson1993},
                                  (8) \citet{Hewitt1993},
                                  (9) \citet{vanHaarlem1993},
                                  (10) \citet{Bershady1994},
                                  (11) \citet{Darling1994},
                                  (12) \citet{Trevese1994},
                                  (13) \citet{Borra1996},
                                  (14) \citet{Willmer1996},
                                  (15) \citet{Veron-Cetty1996},
                                  (16) \citet{Munn1997},
                                  (17) \citet{Bershady1998},
                                  (18) \citet{Treyer1998},
                                  (19) \citet{Ledoux1999},
                                  (20) \citet{Sullivan2000},
                                  (21) \citet{Salzer2001},
                                  (22) \citet{Castander2001},
                                  (23) \citet{Moore2002},
                                  (24) \citet{Wegner2003},
                                  (25) \citet{Jangren2005},
                                  (26) \citet{Trevese2007},
                                  (27) \citet{Hewett2010},
                                  (28) \citet{Edwards2011},
                                  (29) \citet{Hakobyan2012},
                                  (30) \citet{Takey2013},
                                  (31) \citet{Bilicki2014},
                                  (32) \citet{Wen2015},
                                  (33) \citet{Ann2015},
                                  (34) \citet{Rines2016},
                                  (35) \citet{Lansbury2017},
                                  (36) \citet{Haynes2018},
                                  (37) \citet{Ruiz-Lara2018},
                                  (38) \citet{Chilingarian2019},
                                  (39) \citet{Yao2019},
                                  (40) \citet{Lal2020},
                                  (41) \citet{Healy2021},
                                  (42) \citet{Saifollahi2022},
                                  (43) \citet{Liu2023},
                                  (44) \citet{Zaritsky2023}
                                  }
                \tablenotetext{c}{(0) Non-member, (1) Member of the Coma cluster.}
        \end{deluxetable*}
    
        We conducted spectroscopic observations using the 6.5m MMT telescope with the Hectospec \citep{Fabricant2005} to obtain redshifts of galaxies in the field of the Coma cluster.
        These observations are part of the KIAS redshift survey of nearby galaxy clusters.
        Hectospec is a 300-fiber multiobject spectrograph designed for use with the MMT telescope.
        Hectospec provides a spectral resolution of $R\sim 1000 - 2000$ in the wavelength range $3650\mathrm{\AA} - 9500\mathrm{\AA}$.
        With these instruments, we observed 12 fields for 11 nights and obtained 3048 spectra of galaxies.
        Information about the observed fields is summarized in Table \ref{tab:obs_field}.
        Figure \ref{fig:fov} shows the MMT survey region covered in this study (cyan box), overlayed on top of the weak-lensing convergence ($\kappa$) map from \citet{Okabe2014} (left) and the \textit{ROSAT} (\citealp{Truemper1982}) X-ray map (right).

        For the extraction of one-dimensional spectra, we use \texttt{HSRED}, an IDL reduction pipeline for the MMT instruments.
        We then measure the redshifts of each spectrum using \texttt{RVSNUpy} (T. Kim et al. in preparation).
        \texttt{RVSNUpy} is a Python package for measuring the redshifts of spectra by cross-correlating them with known template spectra.
        \texttt{RVSNUpy} provides the Tonry \& Davis $r_{\mathrm{TD}}$ value \citep{Tonry1979} for each redshift measurement, which can be used as a measure of reliability.
        \citet{Geller2014ApJS, Geller2016ApJS} suggested by visual inspection that redshifts with $r_{TD} > 4$ are reliable.
        We use redshifts with \textcolor{black}{$r_{TD} > 4.5$} to be conservative, which is the same criterion used by \citet{Kang2024_Caustic}.
        As a result, we have \textcolor{black}{2184} galaxies with reliable redshifts from the MMT/Hectospec observations.
        In addition, among the 12 point sources observed with the MMT/Hectospec in this study, the redshifts of 10 objects were measured reliably.
        However, all 10 objects turn out to be non-members of the Coma cluster (see Section \ref{sec:data:caustic} for cluster membership determination).

        \begin{figure*}
            \centering
            \includegraphics[width=0.55\linewidth]{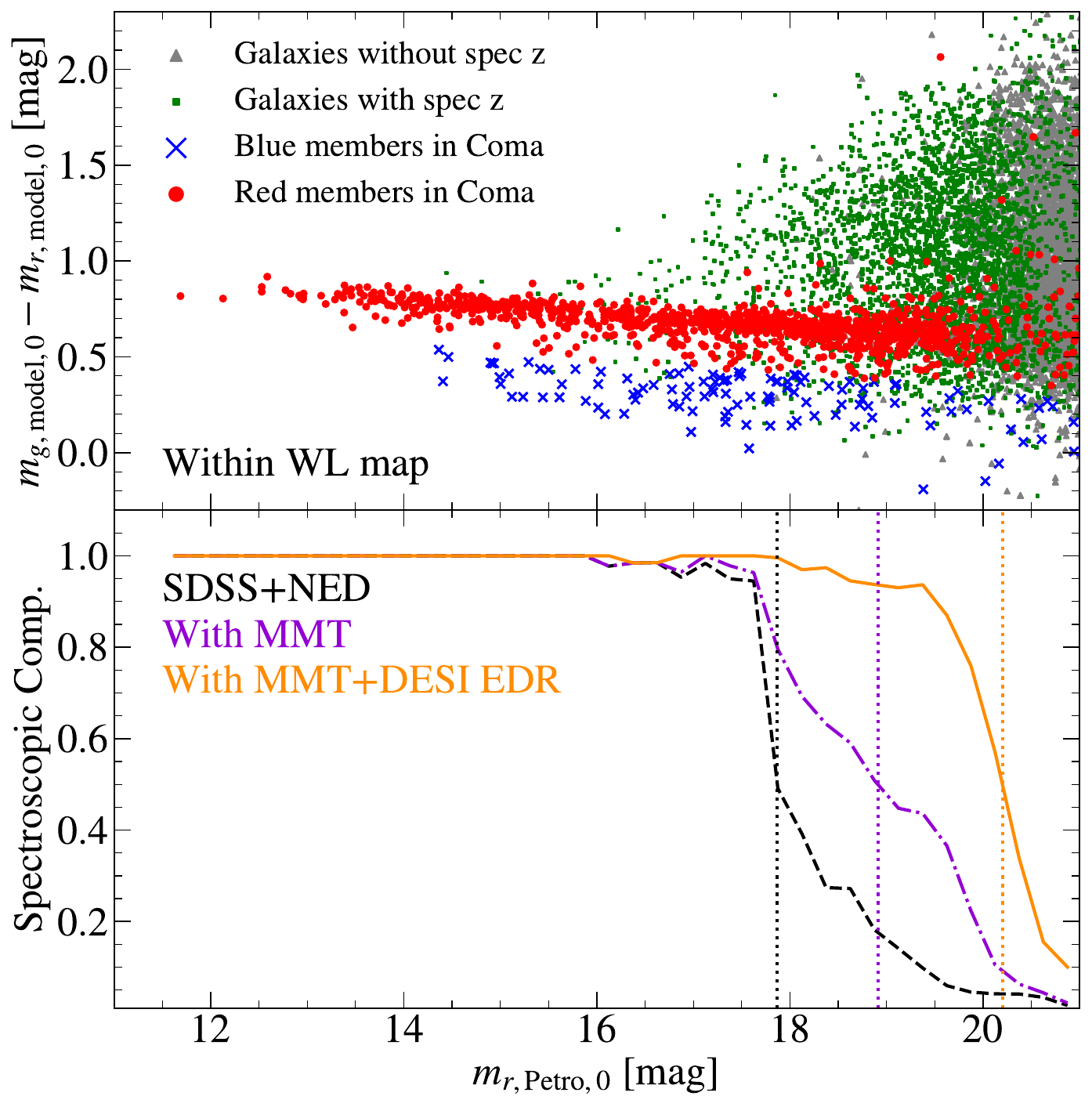}
            \caption{
                Top: Color-magnitude diagram for galaxies in the weak-lensing map.
                Gray triangles, green squares, blue crosses, and red circles represent galaxies without spectroscopic redshift, galaxies with spectroscopic redshift but not members of the Coma cluster, blue members, and red members, respectively.
                For clarity, only 50\% of the gray triangles and green squares are shown.
                Bottom: Differential spectroscopic completeness for galaxies in the weak-lensing map of the Coma cluster.
                The black dashed, purple dot-dashed, and orange solid lines represent spectroscopic completeness of redshift data from SDSS and NED, those combined with the redshifts measured with our MMT/Hectospec observations, and all data available, respectively. 
                The vertical dotted lines indicate the magnitude limits at which the spectroscopic completeness drops below 50\%.
                }
            \label{fig:cmd}
        \end{figure*}

        To fully analyze the 4.5 deg$^2$ region\footnote{The area is wider than the actual region of the weak-lensing map (4.1 deg$^2$) due to the masked regions.} covered by the weak-lensing analysis of \citet{Okabe2014} shown in the left panel of Figure \ref{fig:fov}, we also compile spectroscopic redshifts of galaxies available in the literature.
        Within the weak-lensing field, we retrieve \textcolor{black}{2257} galaxy redshifts from SDSS DR17 and \textcolor{black}{73} from NASA/IPAC Extragalactic Database (NED).
        In addition, the Early Data Release of the Dark Energy Spectroscopic Instrument (DESI EDR; \citealp{DESIEDR2024}) became available in 2023.
        DESI is a wide spectroscopic redshift survey program and its Early Data Release covers the Coma cluster region.
        We include \textcolor{black}{8477} redshifts from DESI EDR for galaxies in the weak-lensing map region without previously measured redshifts in this field.
        The total number of galaxies with measured redshifts is \textcolor{black}{12990} in the 4.5 deg$^2$ field.
        In Table \ref{tab:catalog}, we present the compiled catalog including objects in a wider field ($R=132\arcmin$) which we use for determining cluster membership (see Section \ref{sec:data:caustic} for cluster membership identification).
        The catalog includes SDSS ObjID, coordinates, extinction-corrected $r$-band Petrosian apparent magnitude $m_{r,\mathrm{Petro},0}$, redshift $z$, redshift sources, and cluster membership of galaxies in the Coma cluster.

        We plot the color-magnitude diagram (CMD) in the top panel of Figure \ref{fig:cmd}.
        We use $m_{r,\rm{Petro,0}}$ for the magnitude and the difference between the extinction-corrected model $g$-band apparent magnitude $m_{g,\rm{model,0}}$ and $r$-band magnitude $m_{r,\rm{model,0}}$ for the color.
        The member galaxies of the Coma cluster form a clear red sequence in the CMD.
        We fit the red sequence to a linear function and obtain \textcolor{black}{$-0.034 m_{r,\rm{Petro,0}} + 1.253$} with a scatter \textcolor{black}{$\sigma=0.075$} around the red sequence.
        Following \citet{Hwang2014}, we divide the member galaxies into red members and blue members.
        Member galaxies that are bluer than the $3\sigma$ scatter from the red sequence are classified as blue members, and the others are classified as red members.
        As a result, \textcolor{black}{1337} member galaxies within the weak-lensing map are divided into \textcolor{black}{1209} red members and \textcolor{black}{128} blue members.

        \begin{figure}[!tb]
            \centering
            \includegraphics[width=1.0\linewidth]{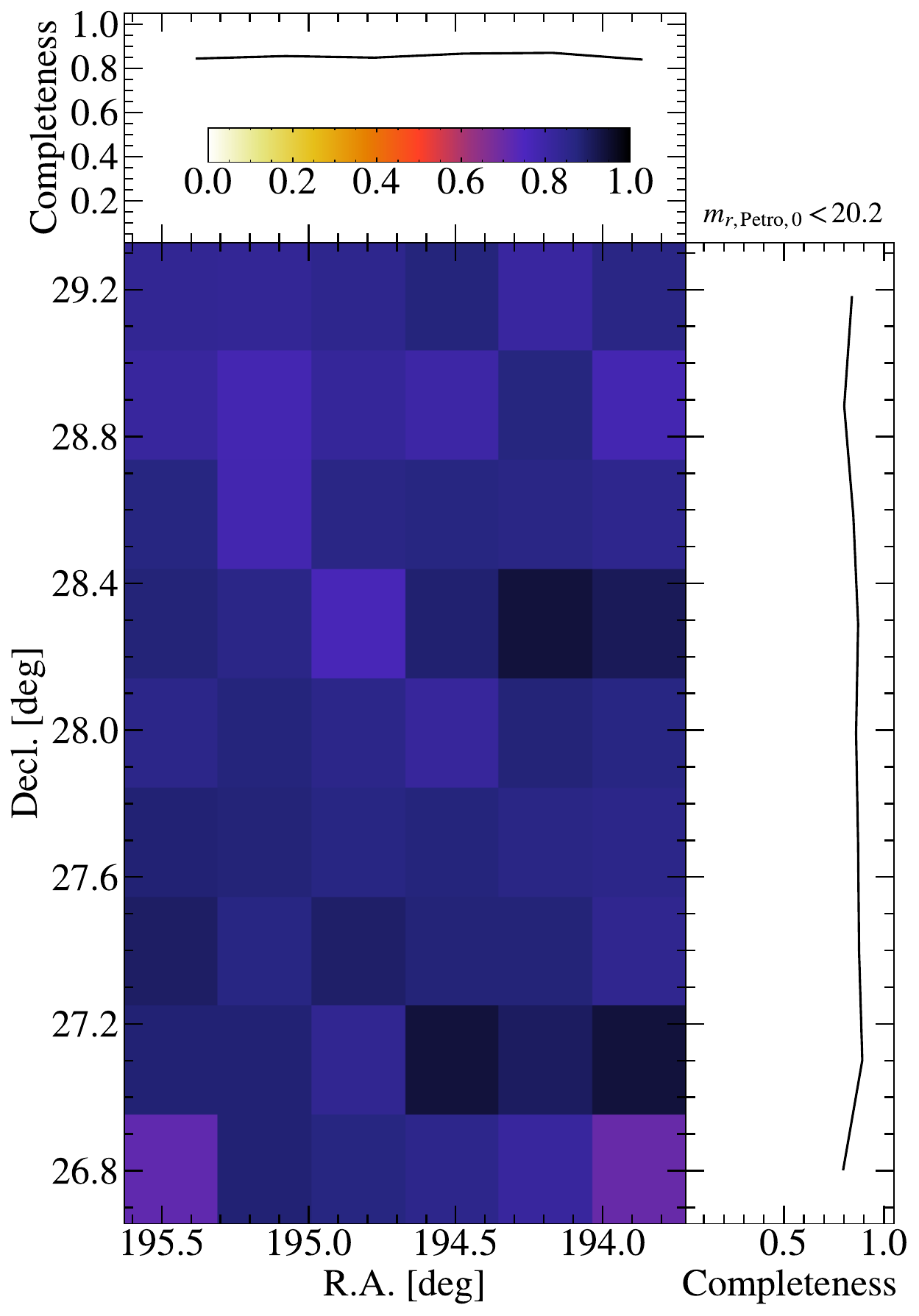}
            \caption{
                Two-dimensional spectroscopic completeness in the field of the Coma cluster weak-lensing map as a function of right ascension and of declination.
                Marginal spectroscopic completenesses as functions of right ascension (top left) and declination (bottom right) are also shown.
            \label{fig:2dcomp}}
        \end{figure}

        \begin{figure}[!htb]
            \centering
            \includegraphics[width=1.0\columnwidth]{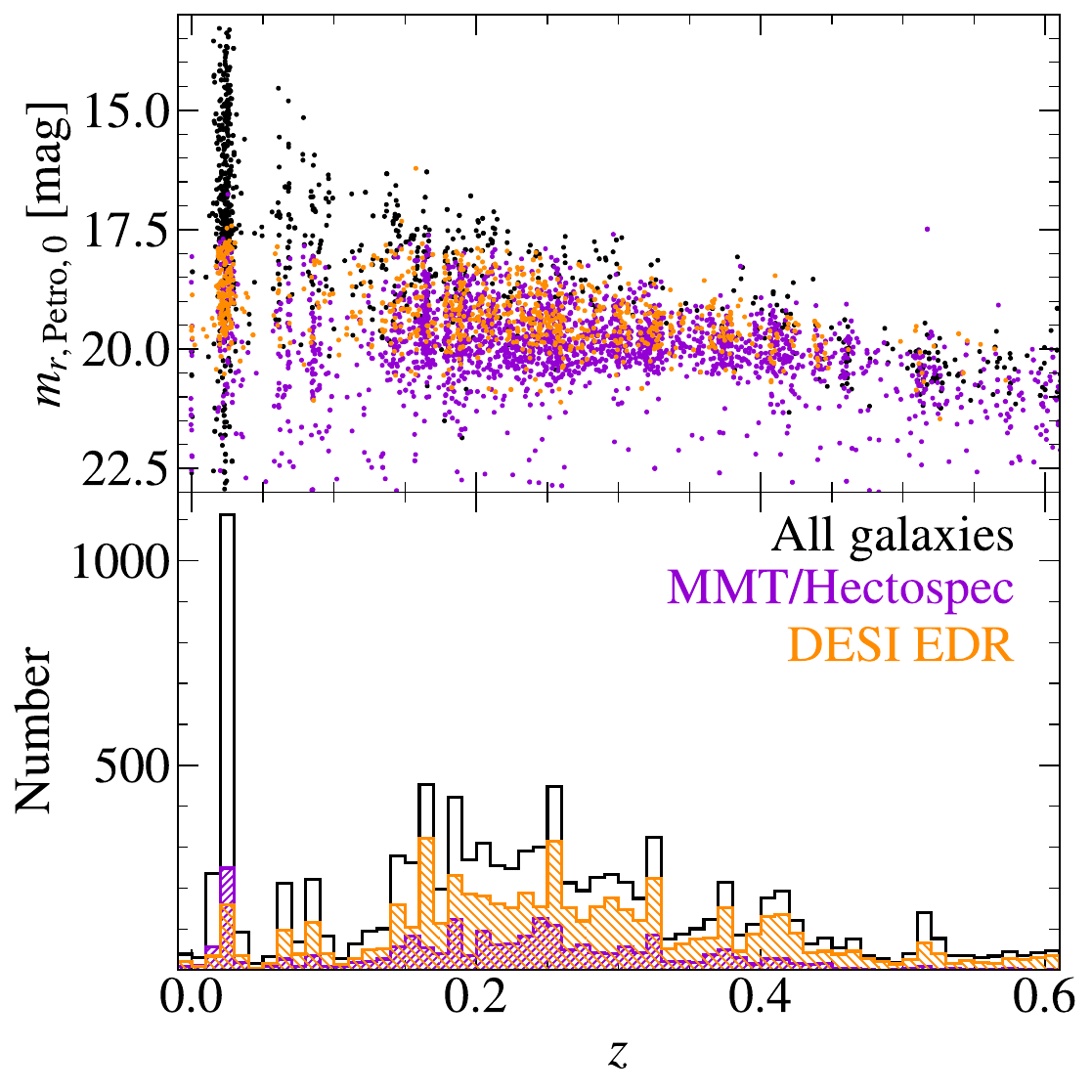}
            \caption{
                Physical parameters of galaxies in the weak-lensing map as a function of redshift.
                Top: Extinction-corrected apparent $r$-band magnitude.
                Black, purple, and orange dots are galaxies with measured redshifts from SDSS and NED, this study, and DESI EDR, respectively.
                We show 50\% of the data points for clarity.
                Bottom: Redshift histogram.
                The blue and red histograms represent data obtained with our MMT/Hectospec observations and retrieved from DESI EDR, respectively.
                The black histogram shows the redshift distribution of all galaxies.
            }
            \label{fig:phys-z}
        \end{figure}
    
        The bottom panel of Figure \ref{fig:cmd} shows the differential spectroscopic completeness of the compiled galaxy redshift data as a function of $m_{r,\rm{Petro,0}}$.
        The spectroscopic completeness is calculated as the ratio of the number of galaxies with spectroscopic redshifts to the number of total galaxies in each magnitude bin.
        We compare the spectroscopic completeness of the redshift data from the SDSS and NED (black dashed line), all that combined with the MMT/Hectospec data from this study (purple dot-dashed line), and with all the available redshift data including the DESI EDR (orange solid line).
        The vertical dotted lines indicate the magnitude at which the differential spectroscopic completeness drops below 50\%. Before this study and DESI EDR, the 50\% magnitude limit was \textcolor{black}{17.9 mag}; including the MMT/Hectospec data from this study improves the 50\% magnitude limit to \textcolor{black}{18.9 mag}; including the DESI EDR redshifts pushes the limit to \textcolor{black}{20.2 mag}.
        The cumulative completeness for galaxies brighter than \textcolor{black}{20.2 mag} is \textcolor{black}{85\%}.

        In Figure \ref{fig:2dcomp}, we show the two-dimensional (2D) spectroscopic completeness as a function of right ascension and declination in the weak-lensing map.
        The spectroscopic completeness is calculated for galaxies brighter than \textcolor{black}{20.2 mag}.
        The top and bottom right panels of Figure \ref{fig:2dcomp} show the spectroscopic completeness marginalized along the declination and the right ascension, respectively.
        The spectroscopic completeness remains above 80\% for most of the regions.
        This figure, together with the bottom panel of Figure \ref{fig:cmd}, shows that the spectroscopic completeness of the data is high and spatially uniform, making them suitable for examining the spatial and kinematic properties of the galaxies in the Coma cluster.
    
        Figure \ref{fig:phys-z} shows the distribution of $m_{r,\rm{Petro,0}}$ (top panel) and the number of galaxies (bottom panel) as a function of redshift.
        From the top panel, we can see that the redshifts obtained from the MMT/Hectospec survey (blue dots) complement the galaxy data at fainter magnitudes, consistent with Figure \ref{fig:cmd}.
        It is also evident that most galaxies are located at the redshift of the Coma cluster ($z=0.0230$), while there still exists a non-negligible number of galaxies in the background.
        
    \subsection{Cluster Member Identification}\label{sec:data:caustic}
        \begin{figure*}[htb]
            \centerline{\includegraphics[width=0.60\linewidth]{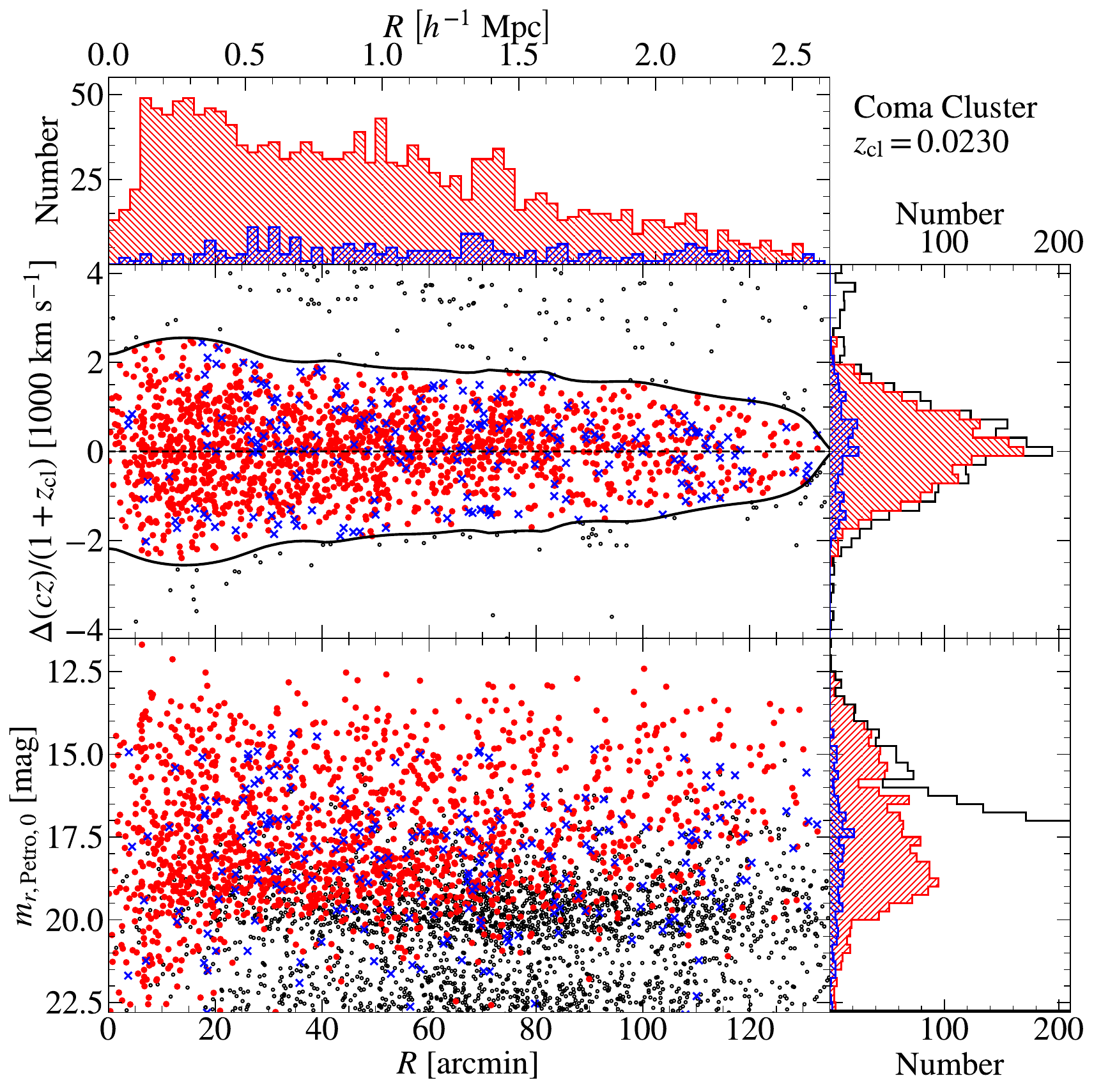}}
            \caption{
                Distribution of galaxies in the redshift space and along the apparent magnitude ($m_{\mathrm{r,Petro},0}$).
                Red filled circles, blue crosses, and black open circles represent red members, blue members, and non-members, respectively (middle left and bottom left panels).
                Red and blue hatched histograms show the marginalized histograms of red and blue galaxies along the clustercentric distance (top left), line-of-sight velocity (middle right), and $m_{\mathrm{r,Petro},0}$ (bottom right).
                Black histograms show distribution of all galaxies for the middle left and bottom right panels.
            \label{fig:caustic}}
        \end{figure*}
        
        It is important to correctly identify the member galaxies of the Coma cluster in order to minimize the contamination from background galaxies.
        One method is to use the caustic technique \citep{Diaferio1999}.
        It is known that cluster member galaxies form a trumpet-shaped distribution in the redshift space (i.e., the plane drawn by the line-of-sight velocity as the ordinate and the projected clustercentric distance as the abscissa; middle left panel of Figure \ref{fig:caustic}). 
        The boundary of this distribution is called the caustics \citep{Kaiser1987, Regos1989, Diaferio1997}.
        The amplitude of the caustics is related to the escape velocity at a given radius. 
        \citet{Diaferio1999} details on how to find the caustics given the right ascension, declination, and redshift of each galaxy.
        \citet{Serra2013} showed, using galaxy clusters created by numerical simulation, that identifying galaxies within the caustics recovers 95\% of the true members and contains 8\% of interlopers within $3R_{200}$.
        Here, $R_{200}$ is the radius within which the mean density is equal to 200 times the critical density of the universe.
        Thus, the caustic technique is a reliable method for separating cluster members from background and foreground galaxies.

        We use \texttt{CausticSNUpy} \citep{Kang2024_Caustic} to determine the cluster membership. 
        \texttt{CausticSNUpy} is a Python package that implements the caustic technique as described by \citet{Diaferio1999} and \citet{Serra2011}.
        \texttt{CausticSNUpy} identifies cluster members as follows.
        The code first finds candidate members based on the pairwise binding energy between all galaxies given as input.
        Then, the number density $f(R, v)$ of all galaxies in the redshift space is estimated via adaptive kernel density estimation.
        The location of the caustics is determined by the contour $f(R,v)=\kappa$, where $\kappa$ is a certain threshold.
        The threshold $\kappa$ is chosen such that 
        \begin{equation}
            S(\kappa)=(\langle v_{\rm{esc}}^{2} \rangle_{\kappa} -4\langle v^{2}\rangle)^{2}
        \end{equation} is minimized.
        Here, $v_{\rm{esc}}$ is the escape velocity of the cluster estimated from the caustics for a given $\kappa$, and $\langle v^{2}\rangle$ is the mean squared velocity of the candidate members found in the first step.
        For a dynamically relaxed cluster, member galaxies will lie in side the contour satisfying $S(\kappa)=0$.
        By minimizing $S(\kappa)$ we find the location of the caustics regardless of the dynamical state of the cluster.
        Finally, the cluster members are determined as the galaxies within the caustics.
        A detailed description of the procedure can be found in \citet{Diaferio1999, Serra2011, Kang2024_Caustic}.
            
        For the membership determination, we also use redshift data in a wider field ($R=132\arcmin$) from the literature.
        We note that the member galaxies of the Coma cluster extends out to $\gtrsim 5h^{-1}\mathrm{Mpc}$ or $250\arcmin$ (see \citealp{Rines2013} for an example).
        Thus, the result of the caustic technique at the cluster outskirts ($R>120\arcmin$) would require a galaxy redshift catalog in a wider region.
        The weak-lensing subhalos studied here are limited to $<90\arcmin$ and the current sample of galaxies is sufficient for our aim.
        When calculating the caustics, we limit the velocity range to $\pm4500\mathrm{km\ s^{-1}}$ from the cluster redshift.
        The resulting caustic lines in the redshift space are shown in the middle left panel of Figure \ref{fig:caustic}.
        We identify a total of \textcolor{black}{1826} galaxies within the caustics as the cluster members, of which \textcolor{black}{1337} galaxies are within the weak-lensing map.
        The bottom left panel of Figure \ref{fig:caustic} shows the distribution of $m_{r, \rm{Petro}, 0}$ along the projected clustercentric distance.
        In the middle left and bottom left panels of Figure \ref{fig:caustic}, open black circles, filled red circles, and blue crosses represent non-members with spectroscopic redshifts, red members, and blue members, respectively.
        The top left, middle right, and bottom right panels of Figure \ref{fig:caustic} are the histograms of galaxies along the projected distance from the cluster center, line-of-sight velocity, and $m_{r, \rm{Petro}, 0}$.
        Red and blue hatched histograms are for red and blue members, respectively, while black open histograms are for all galaxies with spectroscopic redshifts.

        \begin{figure*}[!htb]
            \centerline{\includegraphics[width=0.75\linewidth]{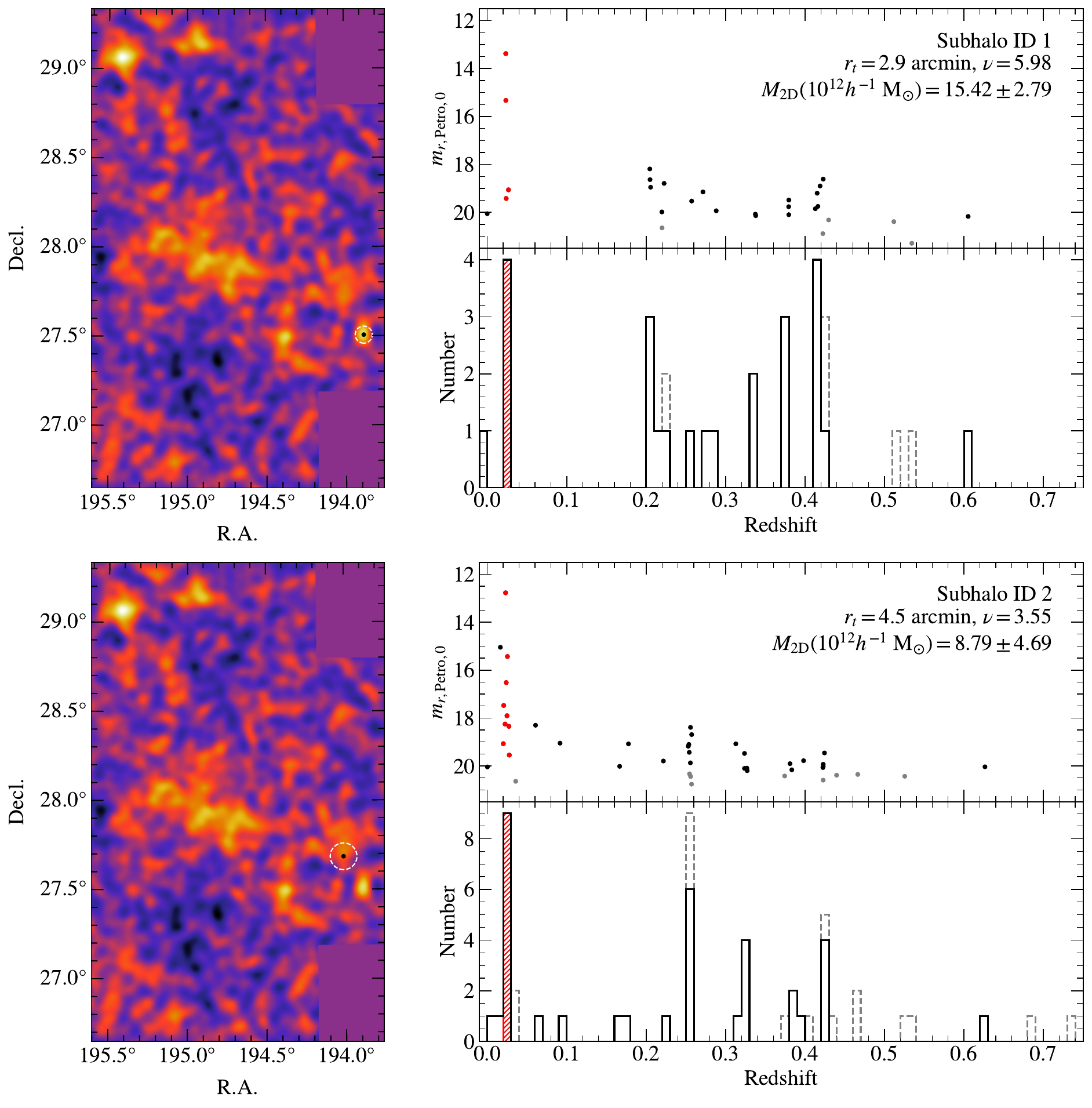}}
            \caption{
                Redshift distribution of galaxies within $r_{t}$ of each subhalo.
                Left: The subhalo center is marked on top of the weak-lensing convergence map as a black dot and the radius of $r_{t}$ is indicated by the white dashed circle.
                Top right: Red and black dots indicate cluster member galaxies and non-members brighter than 20.2 mag, respectively.
                Galaxies fainter than 20.2 mag are shown as pink dots (member galaxies) and gray dots (non-members).
                Bottom right: Red and black histograms show the distribution of member galaxies and all galaxies brighter than 20.2 mag with observed spectroscopic redshifts. 
                The gray dashed-histogram represents the total distribution of galaxies including galaxies fainter than 20.2 mag.
                The complete figure set (32 images) is available in the online journal.
            }
            \label{fig:subhalo-zhist}
        \end{figure*}

        \texttt{CausticSNUpy} also provides the cluster center determined from the candidate cluster members and the mass profile of the cluster according to the methods described by \citet{Serra2011}.
        The cluster center calculated with \texttt{CausticSNUpy} is \textcolor{black}{$(\alpha, \delta)=(195\fdg087, 28\fdg070)$} with a redshift \textcolor{black}{$z_{\mathrm{cl}}=0.0230$}.
        This cluster center has angular separations of \textcolor{black}{11\farcm98} from NGC 4874 and \textcolor{black}{7\farcm23} from NGC 4889, both of which are the central dominant galaxies of the Coma cluster.
        The redshift of the cluster calculated with \texttt{CausticSNUpy} is consistent with that from \citet{Sohn2017ApJS}.
        From the mass profile, we calculate $R_{200}$ and $M_{200}$ (i.e., the enclosed mass within $R_{200}$) of the Coma cluster to be \textcolor{black}{$R_{200}=1.71h^{-1}\ \mathrm{Mpc}$} and \textcolor{black}{$M_{200}=(1.18\pm0.20) \times 10^{15} h^{-1} \mathrm{M_{\Sun}}$}.
        Here, the uncertainty of $M_{200}$ empirically corresponds to the 50\% confidence interval \citep{Serra2011}.
        Our mass measurement is consistent within $2\sigma$ with previous mass measurements of the Coma cluster using weak-lensing analysis ($6.23^{+2.53}_{-1.58} \times 10^{14}h^{-1} \mathrm{M_{\Sun}}$; \citealp{Okabe2014}) and the caustic method ($9.03^{+1.05}_{-1.05} \times 10^{14} h^{-1} \mathrm{M_{\Sun}}$; \citealp{Sohn2017ApJS}).
        Both $M_{200}$ and $R_{200}$ from this study are also in good agreement with the values estimated by \citet{Ho2022}, who used deep learning technique to infer the cluster mass from the distribution of galaxies in the redshift space: $M_{200}=1.26^{+0.52}_{-0.37} \times 10^{15}h^{-1} \mathrm{M_{\Sun}}$ and $R_{200}=1.78\pm0.03h^{-1}\mathrm{Mpc}$.

    \subsection{Weak-lensing Data}

        \citet{Okabe2014} analyzed deep Subaru/Suprime-Cam images covering a 4.1 deg$^2$ region of the Coma cluster for weak gravitational lensing study.
        They detected 32 subhalos in the Coma cluster and measured their physical properties via weak-lensing analysis.
        These properties include the position of the subhalo center, 2D projected mass $M_{\mathrm{2D}}$, and the truncation radius $r_{t}$.
        $M_{\mathrm{2D}}$ and $r_{t}$ are measured using a model-independent aperture densitometry method \citep{Clowe2000ApJ}, which measures the enclosed projected mass using only the lensing signal within a given radius.
        This allows us to exclude the contribution from surrounding unrelated lensing signals.
        Due to the tidal interaction with the cluster gravitational potential, the mass densities of the subhalos have sharp cutoffs.
        The truncation radius $r_{t}$ is where the enclosed mass profile of the subhalo saturates.

        To take into account the effect of lensing signal due to background large-scale structure (LSS), \citet{Okabe2014} measured the galaxy--galaxy lensing signal and excluded the LSS contribution from the measured shear.
        The LSS lensing signal does not model well the contribution from background galaxy clusters or groups.
        \citet{Okabe2014} also identified subhalos with known background objects (Subhalo IDs 1 and 32).
        For these two subhalos, they fitted the tangential distortion profiles to a combined model of a Navarro-Frenk-White profile (NFW; \citealp{Navarro1996ApJ, Navarro1997ApJ}) for the background object and a truncated NFW profile for the subhalo.
        When the background object is considered, the best-fit masses of the subhalos are decreased.
        The results are presented in Table 5 of \citet{Okabe2014}, which we use in this study.
        
        In Figure \ref{fig:subhalo-zhist}, we show the distribution of galaxies in the line-of-sight of the subhalos.
        The left panels indicates the location and the size $r_{t}$ of the subhalos with the weak-lensing $\kappa$ map in the background.
        The top right panels show the apparent magnitudes $m_{r,\mathrm{Petro},0}$ of galaxies within $r_{t}$ of the subhalo center.
        The bottom panels show the histograms of galaxies in the line of sight, with the cluster member galaxies indicated as the red hatched histogram.
        This figure illustrates that the spectroscopic data can be used to study individual subhalo properties obtained with weak-lensing analysis.
         
    \section{RESULTS} \label{sec:result}
            
        \subsection{The Effect of Background Structures on the Weak-lensing Masses of Subhalos} \label{sec:res:bkg}
               
\begin{figure}
                \centerline{\includegraphics[width=0.83\linewidth]{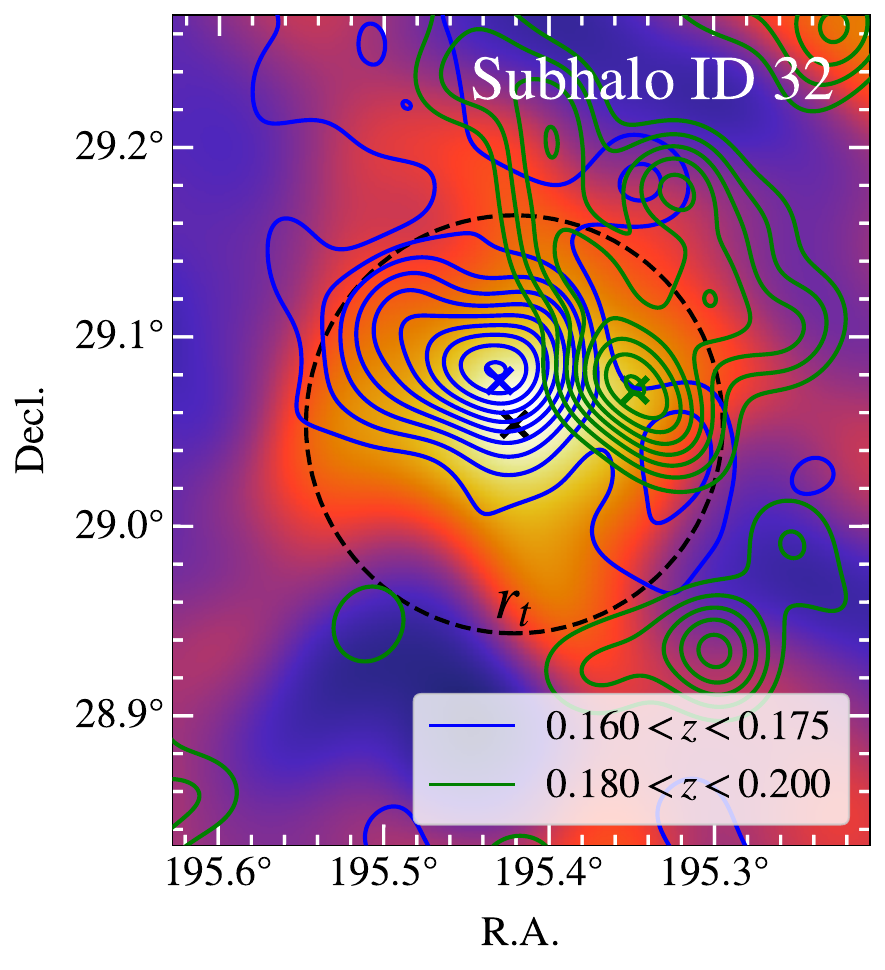}}
                \caption{
                    Background galaxy distributions in the line of sight of Subhalo ID 32.
                    The color image represents the $\kappa$ map from \citet{Okabe2014}.
                    The blue and green contours show the number density of galaxies at redshift $0.160 < z < 0.175$ and $0.180 < z < 0.200$, respectively.
                    The contour levels start from $1\sigma$ above the mean number density with a step size of $1\sigma$.
                    The black, blue, and green crosses indicate the center of the Subhalo ID 32, RMJ130142.6+290438.5, and GMBCGJ195.34791+29.07201, respectively.
                    The black dashed circle show the truncation radius $r_{t}$ of the Subhalo ID 32.
                }
                \label{fig:sub32_bkg}
            \end{figure}

            It is important to check the effect of fore-/background structures on the weak-lensing analysis because gravitational lensing signal takes into account the total mass along the line of sight.
            \citet{Okabe2014} took the effect of background object GMBCG J195.34791+29.07201 \citep{Hao_2010_ApJS_191_254H} at redshift $z=0.189$ in the line of sight of Subhalo ID 32.
            We find that another galaxy cluster exists in the background: RMJ130142.6+290438.5 from the redMaPPer galaxy cluster catalog \citep{Rykoff2014} at redshift $z=0.163$, whose paper was published later than \citet{Okabe2014}.
            In Figure \ref{fig:sub32_bkg}, we show the distribution of galaxies brighter than 20.2 mag and in redshift range $0.160<z<0.175$ (blue contours) and $0.180<z<0.200$ (green contours) on top of the weak-lensing $\kappa$ map.
            The smoothing kernel of the number density is a Gaussian with a full width at half maximum of 4\arcmin, which is the same as the $\kappa$ map.
            From the mass--richness relation, the mass of the background cluster is estimated to be $M_{200}=(3.401\pm0.421)\times 10^{14} \mathrm{M_{\odot}}$ with $h=0.7$ \citep{Sereno2017}.
            The effect of this background cluster is not considered in the mass measurement of Subhalo ID 32.
            As such, it is possible that the weak-lensing mass of Subhalo ID 32 has been slightly overestimated. 
            We exclude the subhalo from our analysis but show it in the plots of Section \ref{sec:res:phys_prop} for comparison with other subhalos.

             \begin{figure}
                \centerline{\includegraphics[width=0.8\columnwidth]{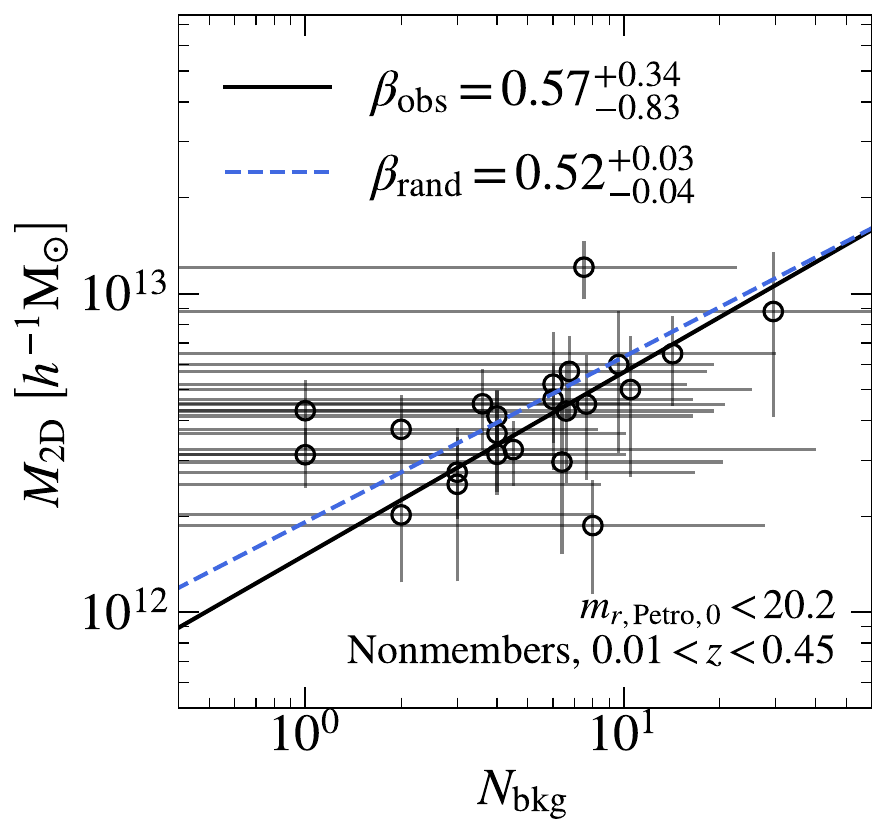}}
                \caption{
                    Relation between $M_{\mathrm{2D}}$ and $N_{\mathrm{bkg}}$, the number of nonmember galaxies in redshift $0.01<z<0.45$ and within $r_{t}$ of each subhalo center.
                    Only nonmember galaxies with $m_{r,\mathrm{Petro},0} < 20.2$ are counted.
                    The black solid line represents the best-fit power law of the observed data, while the blue dashed line shows the power-law relation of random sky pointings using the same aperture set.
                    Note that the error for $\beta_{\mathrm{rand}}$ is the 68\% interval for 1000 random realizations, not the uncertainties of individual realizations.
                }
                \label{fig:bkg_test}
            \end{figure}
            
            We also examine if the weak-lensing mass of the subhalos are influenced by the number of background galaxies in the line of sight.
            In Figure \ref{fig:bkg_test}, we plot the relation between the subhalo mass $M_{\mathrm{2D}}$ and the number of nonmember galaxies within $r_{t}$ of each subhalo center $N_{\mathrm{bkg}}$.
            We only count nonmembers within redshift range $0.01 < z < 0.45$ and with $m_{r,\mathrm{Petro},0} < 20.2$.
            The error of $N_{\mathrm{bkg}}$ consists of the Poisson error $\sqrt{N_{\mathrm{bkg}}}$ and the uncertainty due to the measurement error of $r_{t}$.
            Because $N_{\mathrm{bkg}}\propto r_{t}^{2}$ for a given aperture $r_{t}$, the error $\delta r_{t}$ propagates to the uncertainty in $N_{\mathrm{bkg}}$ as $2\delta r_{t}N_{\mathrm{bkg}}/r_{t}$.
            Thus, the total error in $N_{\mathrm{bkg}}$ is $\sqrt{N_{\mathrm{bkg}} + (2\delta r_{t}N_{\mathrm{bkg}}/r_{t})^{2}}$.
            We exclude Subhalo IDs 1 and 32 from the plot and the analysis of the background galaxy effects, because there are known background structures in the line of sight of Subhalo IDs 1 and 32 and their effects on the mass measurement are understood.
            We fit the data to a power law relation $\ln M_{\mathrm{2D}}/(10^{12}h^{-1}\mathrm{M_{\Sun}}) = \alpha_{\mathrm{obs}} + \beta_{\mathrm{obs}} \ln N_{\mathrm{bkg}}$, where $\ln$ is the natural logarithm.
            
            To properly obtain the best-fitting parameters $\alpha_{\mathrm{obs}}$ and $\beta_{\mathrm{obs}}$, we need to take into account the errors in $M_{\mathrm{2D}}$ and $N_{\mathrm{bkg}}$ and the selection effect.
            In our case, the selection effect is mostly due to the detection limit of the weak-lensing analysis.
            We use a hierarchical Bayesian regression as presented by \citet{Akino_2022PASJ_74_175A} (see also \citealp{Sereno_2016MNRAS_455_2149S}).
            The prior is a uniform distribution between $-10^{4}$ and $10^{4}$ for $\alpha_{\mathrm{obs}}$ and a Student's $t$ distribution with one degree of freedom for $\beta_{\mathrm{obs}}$ so that the slope angle has a uniform distribution.

            The best-fitting parameters are $\alpha_{\mathrm{obs}}=0.41^{+1.58}_{-0.77}$ and $\beta_{\mathrm{obs}}=0.57^{+0.34}_{-0.83}$.
            The power-law relation is shown as the black solid line in Figure \ref{fig:bkg_test}.
            Although the result implies that there is a correlation between $M_{\mathrm{2D}}$ and $N_{\mathrm{bkg}}$, it should be noted that $M_{\mathrm{2D}}$ also scales nearly linearly with $r_{t}$ as described by \citet{Okabe2014}.
            If background galaxies are uniformly distributed in the sky, then one would expect $N_{\mathrm{bkg}} \propto r_{t}^{2} \propto M_{\mathrm{2D}}^2$, which is consistent with our results.
            
            To statistically verify whether the background galaxies are not concentrated in the line of sight of the subhalos, we randomly pick positions in the weak-lensing map (excluding the masked regions) and count the number of background galaxies using the same aperture set.
            Again, we exclude the apertures of Subhalo IDs 1 and 32.
            In addition, we also mask the regions around Subhalo IDs 1 and 32 as $2r_{t} \times 2r_{t}$ squares and do not pick within the masked regions.
            We compute the best-fit power law parameters $\alpha_{\mathrm{rand}}$ and $\beta_{\mathrm{rand}}$ between the subhalo mass and the number of background galaxies in random sky positions.
            For 1000 realizations of random sky positions, the median of the best-fit slopes is $\beta_{\mathrm{rand}}=0.52$ with a 16 percentile and 84 percentile values of 0.48 and 0.55, respectively.
            We plot the scaling relations using the median values of the parameters as the blue dashed line in Figure \ref{fig:bkg_test}.
            Considering that the slopes of individual realizations  have uncertainties similar to those of $\alpha_{\mathrm{obs}}$ and $\beta_{\mathrm{obs}}$, the observed scaling relation does not show a significant difference from a random background distribution.
            This result supports the idea that the correlation between $M_{\mathrm{2D}}$ and $N_{\mathrm{bkg}}$ could be mainly because of the uniform distribution of background galaxies and the correlation between $M_{\mathrm{2D}}$ and $r_{t}$.
                
        \subsection{Overcoming Intracluster Projection Effect Using Gaussian Mixture Model} \label{sec:res:gmm}

            \begin{figure*}
                \centerline{\includegraphics[width=0.85\linewidth]{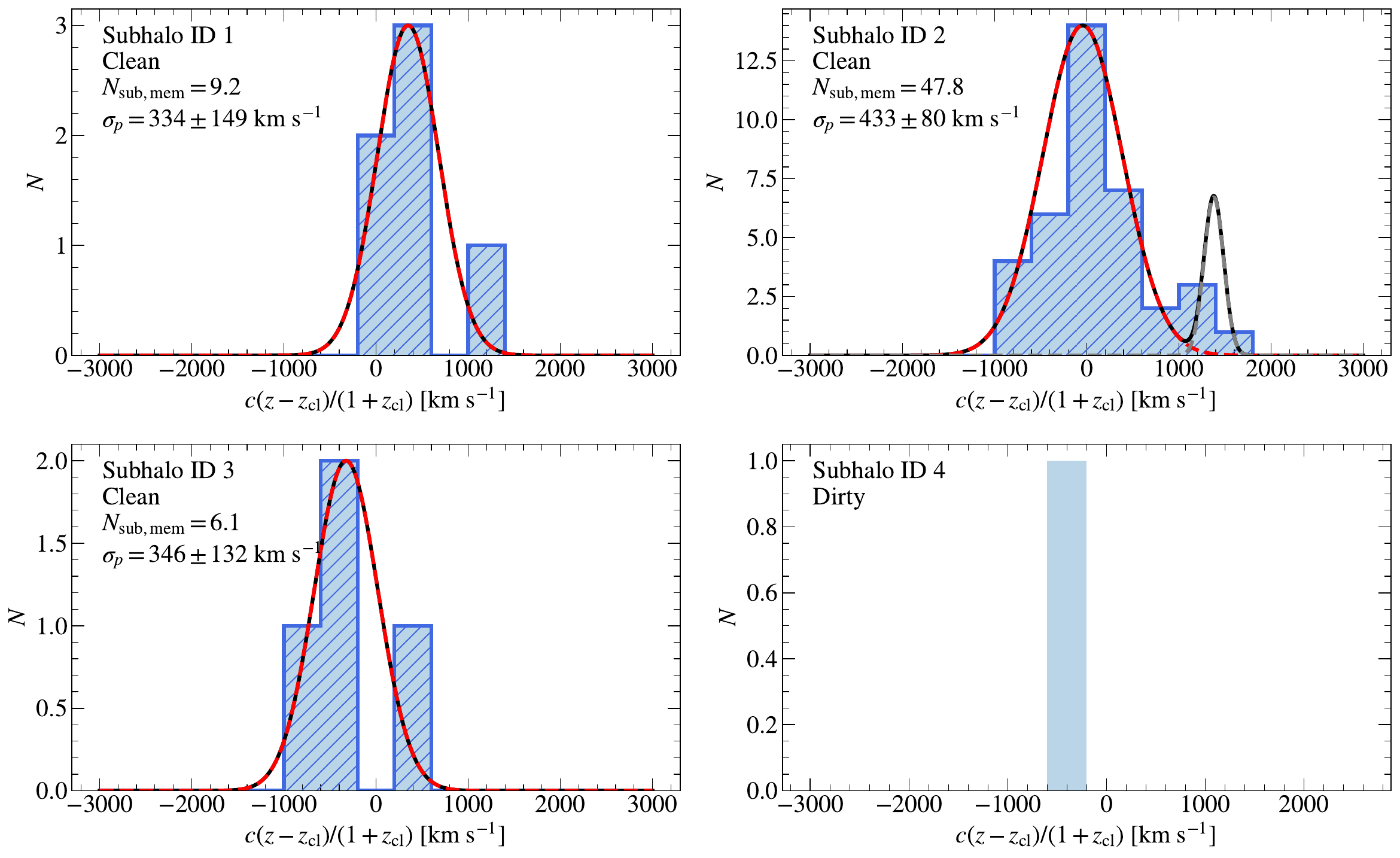}}
                \caption{
                    Gaussian Mixture Modeling for Subhalo IDs 1--4.
                    The blue histogram represents line-of-sight velocities of the Coma members within $2r_{t}$ of the subhalo center.
                    The black solid line shows the sum of the modeled Gaussians.
                    The red dashed line represents the Gaussian distribution of the main component, while the gray dashed lines show the Gaussian distributions of other components.
                    The hatched histogram indicates the galaxies used for calculating the resampling errors of the line-of-sight velocity dispersion.
                    The complete figure set (32 images) is available in the online journal.
                }
                \label{fig:gmm}
            \end{figure*}

            Identifying galaxies associated with individual subhalos and separating them from line-of-sight interlopers is critical in studying the physical nature of the subhalos.
            For example, if one uses all the cluster member galaxies in the line of sight of subhalos, the velocity dispersion would be overestimated due to the unrelated cluster galaxies.
            Therefore, we need to have a way to separate the galaxies within and outside the subhalos in cluster.

            If we assume that the galaxies associated with a subhalo follow the bulk motion of the subhalo itself, the subhalo galaxies will make a peak in the velocity histogram.
            Based on this idea, we propose using Gaussian Mixture Model (GMM) in the line-of-sight velocity domain to measure the physical properties of subhalos, namely, the mean velocity, velocity dispersion, and number of subhalo galaxies.
            As such, line-of-sight substructures moving at different velocities would show up as different peaks in the velocity histogram.
            GMM models the distribution of the data as the sum of finite number of Gaussian distributions.
            It uses expectation-maximization to estimate the parameters of $N$ Gaussian distributions $\mu_{i}$, $\sigma_{i}$, and $w_{i}$, which are the mean, scale, and weight of the $i$-th ($1 \leq i \leq N$) Gaussian component.
            The weight $w_{i}$ is normalized such that the sum is equal to unity and is equal to the probability a data point belonging to the $i$-th component.
            In other words, the product of the number of the total sample and $w_{i}$ will give the expected number of samples drawn from the $i$-th Gaussian component.

            We apply GMM to the velocities of Coma member galaxies within $2r_{t}$, instead of $r_{t}$, of each subhalo center.
            This is to secure a sufficient amount of data points to identify different distributions.
            The choice of $2r_{t}$ also allows the method to be less sensitive to the position accuracy of the subhalo centers and the measurement error of $r_{t}$.
            The effect of choosing a different aperture size is described in the following paragraphs.
            We do not apply the magnitude cut of 20.2 mag as the kinematics within the subhalo is less affected by spectroscopic completeness, contrary to when studying the spatial distributions of galaxies.
            We choose the number of components to model by inspecting the velocity histogram and chaning bin sizes from $300 \ \mathrm{km \ s^{-1}}$ to $450 \ \mathrm{km \ s^{-1}}$.
            The typical escape velocity of the weak-lensing subhalos at $r_{t}$ from the center is $1000 \ \mathrm{km \ s^{-1}}$, which is the ground for choosing the bin sizes.
            
            In Figure \ref{fig:gmm}, we show the distribution of cluster member galaxies within $2r_{t}$ and the result of GMM.
            The black solid solid lines represent the sum of Gaussian probability density.
            We define the main component as the Gaussian component with the highest weight (and thus with the most members) and show the probability density as the red dashed line in Figure \ref{fig:gmm}.
            We use the mean of the main Gaussian component as the line-of-sight bulk velocity ($v_{p}$) of the subhalo.
        
            Because GMM models the sample distribution, the scale of the main Gaussian component is the sample standard deviation and thus is a biased estimator of the true population variance.
            We therefore multiply by $\sqrt{N/N-1}$ to the scale of the main Gaussian component where $N$ is the number of galaxies in the main component, and use this value as the projected velocity dispersion $\sigma_{p}$.
            
            We use the resampling error as the estimate of the error of $\sigma_{p}$.
            For subhalos with clear separation of galaxy velocity distributions (e.g., Subhalo ID 6), it would make sense to resample galaxies only in the main component.
            However, for subhalos modeled as multiple overlapping Gaussian distributions, it is difficult to select galaxies belonging to the main component.
            As a result, we resample galaxies within a given velocity range that encompasses the overlapping Gaussian distributions.
            In Figure \ref{fig:gmm}, we indicate the galaxies used for resampling as the blue hatched histogram.
            We apply GMM to the resampled galaxies and calculate the velocity dispersion of the main component.
            We make 1000 realizations of resampled galaxies for each subhalo and use the $1\sigma$ scatter of the main component's velocity dispersion as the error of $\sigma_{p}$.
                
            We examine possible biases in the estimation of the velocity dispersion.
            First, we test if the standard deviation estimated by GMM is biased in an ideal situation.
            Specifically, we create mock data from a one-component Gaussian and a two-component Gaussian and test whether GMM recovers the true standard deviation.
            For the one-component test, we draw three data points from a Gaussian distribution with a scale of 300 and estimate the standard deviation using GMM.
            For 1000 realizations, the estimated standard deviation is $251^{+150}_{-131}$, where the uncertainty is the 68\% interval.
            We repeat the test for a two-component case, drawing 10 data points from a Gaussian with a mean of 1000 and a scale of 400, and three data points from a Gaussian with a mean of 0 and a scale of 100.
            Again, with 1000 realizations, the estimated standard deviation of the main component is $341^{+115}_{-90}$, where the true standard deviation is 400.
            These tests show that even when two Gaussian components are blended, the bias inherent in the GMM is negligible given the uncertainties of the standard deviation.
            To check for other possible biases in the velocity dispersion at small sample sizes, we reanalyze the results presented in the following subsections using a stricter condition, i.e., using subhalos with five or more subhalo member galaxies.
            However, the results change only within the uncertainties.

            In addition, we examine the effect of using a larger aperture size when applying GMM.
            We apply GMM to cluster member galaxies within $3r_{t}$ of each subhalo and compare the estimated velocity dispersion with those using $2r_{t}$ as the aperture size.
            The velocity dispersion is overall slightly overestimated, typically by $\times 1.3$.
            This is probably due to the inclusion of other unrelated cluster galaxies.
            However, the change in the power-law slope (Section \ref{sec:res:phys_prop}) is negligible considering the uncertainties.

            We investigate whether the location of the caustics would affect the velocity dispersion by clipping the tails of the distributions.
            We use galaxies within the velocity range $\pm 3000 \ \mathrm{km \ s^{-1}}$ regardless of the membership and apply GMM to the velocity distribution.
            The estimated velocity dispersions remain unchanged.
            This is because only one or two galaxies per subhalo are added to the distribution that have a distinct velocity from the main component.        
            
            We calculate the number of subhalo galaxies brighter than 20.2 mag in $r$-band as
            \begin{equation}
                N_{\mathrm{sub,mem}}=w_{\mathrm{main}} N_{\mathrm{cl}, <20.2} \times \frac{N_{\mathrm{total}, <20.2}}{N_{\mathrm{spec}, <20.2}} 
            \end{equation}
            where $w_{\mathrm{main}}$ is the weight of the main component, $N_{\mathrm{cl}, <20.2}$ is the number cluster member galaxies, $N_{\mathrm{total}, <20.2}$ is the number of all galaxies including those without spectroscopic redshifts, and $N_{\mathrm{spec}, <20.2}$ is the number of all galaxies with spectroscopic redshifts.
            All numbers are for galaxies brighter than 20.2 mag and within $2r_{t}$.
            The last term $N_{\mathrm{total}, <20.2}/N_{\mathrm{spec}, <20.2}$ corrects for the spectroscopic incompleteness within the aperture of each subhalo. 
            Thus, $N_{\mathrm{sub,mem}}$ is the magnitude-limited, spectroscopic incompleteness-corrected number of subhalo galaxies which is necessary for comparison between different subhalos.
                        
            Not all velocity distributions show an evident peak, and it is thus necessary to flag the subhalos.
            We flag subhalos as ``clean'' if the number of galaxies in the main component is three or more.
            For example, although Subhalo ID 25 have three galaxies in the line of sight, the main component only has two galaxies and thus is flagged ``dirty.''
            One exception is Subhalo ID 24, which has a sufficient number of galaxies in the main component.
            However, as the velocity distribution is modeled with five closely spaced components, the estimate of the error in the velocity dispersion (elaobrated in the following paragraph) may not be reliable.
            We thus flag Subhalo ID 24 as ``dirty'' to be conservative and have \textcolor{black}{18} ``clean'' subhalos.
            The velocity dispersion and the number of subhalo galaxies for Subhalo ID 24 is $\sigma_{p} = 260 \ \mathrm{km \ s^{-1}}$ and $N_{\mathrm{sub,mem}} = 25.8$.
            The result of our analysis does not change when we include this subhalo.
            In total, we have 18 subhalos flagged as ``clean,'' including Subhalo ID 32 which we do not use for our analyses.
            We summarize the subhalo properties measured with GMM in Table \ref{tab:gmm_res}.

            \begin{deluxetable}{ccccc}
                    \label{tab:gmm_res}
                    \caption{Subhalo properties measured with GMM.}
                    \tablehead{
                        \colhead{Subhalo ID} & \colhead{$v_{p}$} & \colhead{$\sigma_{p}$} & \colhead{$N_{\mathrm{sub,mem}}$} & \colhead{Flag} \\ 
                        \colhead{} & \colhead{$(\mathrm{km \ s^{-1}})$} & \colhead{$(\mathrm{km \ s^{-1}})$} & \colhead{} & \colhead{}
                    }
                    \startdata
                        1\tablenotemark{a}  &   $348$ & $334 \pm 149$ & $ 9.2$ & Clean \\
                        2                   &   $-45$ & $433 \pm  80$ & $47.8$ & Clean \\
                        3                   &  $1325$ & $346 \pm 132$ & $ 6.1$ & Clean \\
                        4                   &   $...$ &      ...      & $ ...$ & Dirty \\
                        5                   &   $289$ & $191 \pm  78$ & $ 6.9$ & Clean \\
                        6                   &   $932$ & $125 \pm  70$ & $ 4.6$ & Clean \\
                        7                   &   $102$ & $151 \pm  65$ & $ 9.8$ & Clean \\
                        8                   &   $...$ &      ...      & $ ...$ & Dirty \\
                        9                   &   $494$ & $409 \pm 124$ & $26.4$ & Clean \\
                        10                  &   $...$ &      ...      & $ ...$ & Dirty \\
                        11                  &   $...$ &      ...      & $ ...$ & Dirty \\
                        12                  &   $...$ &      ...      & $ ...$ & Dirty \\
                        13                  &   $...$ &      ...      & $ ...$ & Dirty \\
                        14                  &   $461$ & $312 \pm 107$ & $ 6.1$ & Clean \\
                        15                  &  $1553$ & $233 \pm  65$ & $ 7.6$ & Clean \\
                        16                  &   $...$ &      ...      & $ ...$ & Dirty \\
                        17                  &   $...$ &      ...      & $ ...$ & Dirty \\
                        18                  &  $-438$ & $487 \pm 127$ & $25.9$ & Clean \\
                        19                  &   $139$ & $210 \pm  76$ & $ 3.8$ & Clean \\
                        20                  &   $280$ & $165 \pm  52$ & $ 7.6$ & Clean \\
                        21                  &   $ 68$ & $145 \pm  64$ & $12.2$ & Clean \\
                        22                  &   $255$ & $297 \pm 125$ & $ 4.6$ & Clean \\
                        23                  &  $1056$ & $324 \pm 111$ & $ 6.5$ & Clean \\
                        24\tablenotemark{b} &   $295$ & $260 \pm  96$ & $25.8$ & Dirty \\
                        25                  &   $...$ &      ...      & $ ...$ & Dirty \\
                        26                  &   $...$ &      ...      & $ ...$ & Dirty \\
                        27                  &   $763$ & $333 \pm  71$ & $10.0$ & Clean \\
                        28                  &   $...$ &      ...      & $ ...$ & Dirty \\
                        29                  &  $-918$ & $200 \pm  81$ & $ 4.5$ & Clean \\
                        30                  &   $...$ &      ...      & $ ...$ & Dirty \\
                        31                  &   $...$ &      ...      & $ ...$ & Dirty \\
                        32\tablenotemark{c} &   $134$ & $274 \pm  61$ & $21.9$ & Clean \\
                    \enddata
                    \tablenotetext{a}{Background object is considered in the weak-lensing mass by \citet{Okabe2014}.}
                    \tablenotetext{b}{Flagged ``dirty'' due to the complex components in the line of sight.}
                    \tablenotetext{c}{Excluded from our analyses due to the additional background object.}
                    
                \end{deluxetable}

            Although GMM cannot determine the membership of individual galaxies, it can still estimate the statistical properties of subhalo galaxies.
            The strength of GMM is evident when there are two or more blended components in the line of sight.
            In this case, conventional outlier rejection methods such as $3\sigma$ clipping would be insufficient.

            \begin{figure*}
                \centerline{\includegraphics[width=0.63\linewidth]{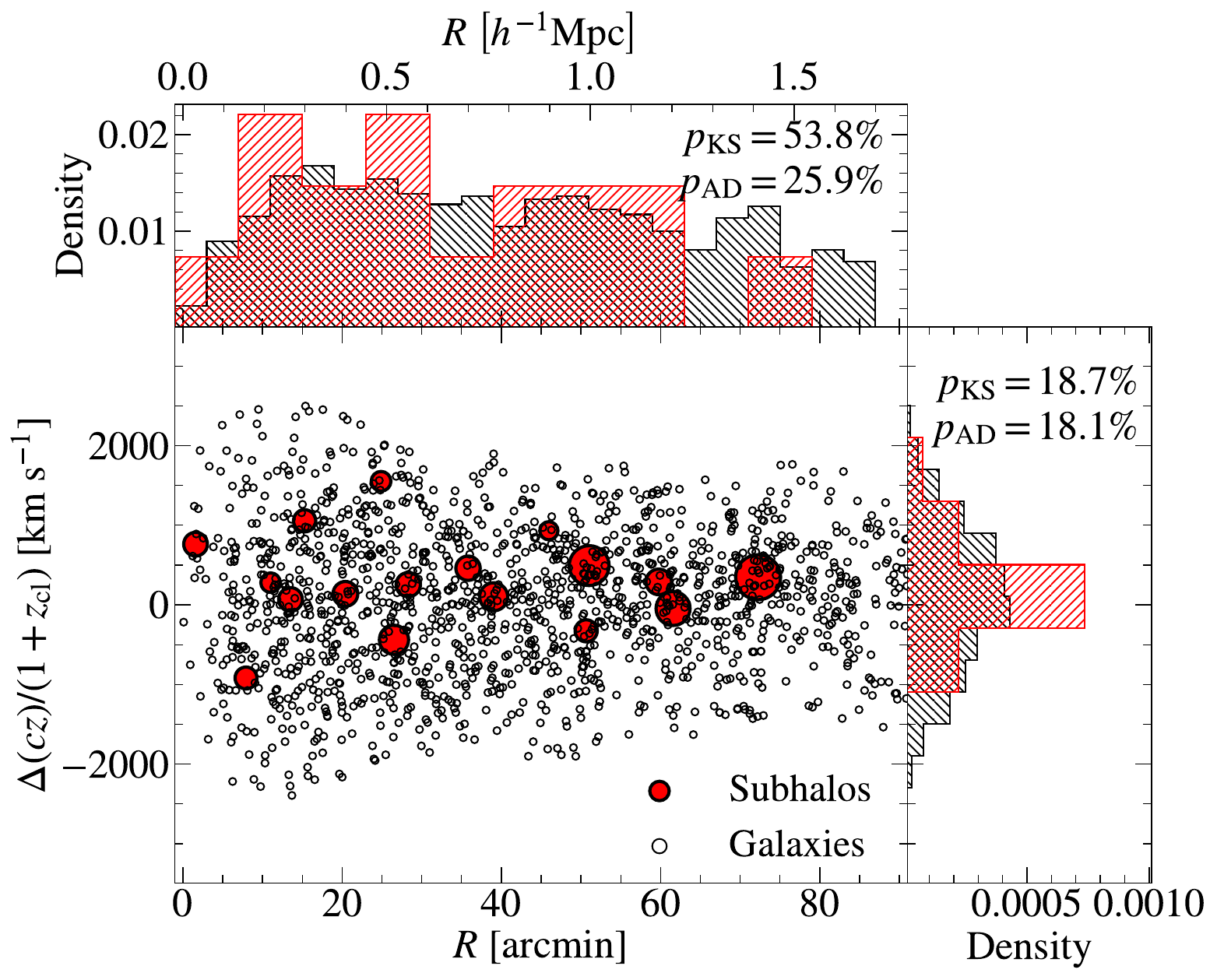}}
                \caption{
                    Bottom left: Line-of-sight velocity as a function of clustercentric radius.
                    Black open circles and red filled circles indicate cluster member galaxies and subhalos, respectively.
                    The size of the circle is proportional to the subhalo mass.
                    Top left and botom right: Histograms of galaxies (black) and subhalos (red).
                }
                \label{fig:subhalo_caustic}
            \end{figure*}
            
        \subsection{Physical Properties of Weak-lensing Subhalos} \label{sec:res:phys_prop}
        
            \subsubsection{Redshift Space Distribution}                
                We examine the distribution of subhalos in the 2D redshift space (Figure \ref{fig:subhalo_caustic}).
                The red circles represent subhalos with their marker radii proportional to $M_{\mathrm{2D}}$, while the black open circles are member galaxies.
                We use the line-of-sight velocities $v_{p}$ of the ``clean'' subhalos determined from GMM (Section \ref{sec:res:gmm}).
                We perform two-sample Kolmogorov-Smirnov (KS) and Anderson-Darling (AD) tests between the distributions of the member galaxies and the subhalos.
                We obtain a $p$-value of \textcolor{black}{$p_{\mathrm{KS}}=53.8\%$} for the KS test and \textcolor{black}{$p_{\mathrm{AD}}=25.9\%$} for the AD test along the clustercentric distance.
                For the line-of-sight velocity distribution, the results are \textcolor{black}{$p_{\mathrm{KS}}=18.7\%$} for the KS test and \textcolor{black}{$p_{\mathrm{AD}}=18.1\%$} for the AD test.
                As a result, we cannot reject the null hypothesis that the subhalos and galaxies are drawn from the same distributions of the clustercentric distance and the line-of-sight velocity.

                We also conduct the KS and AD test for the velocities of the cluster member galaxies and those of the brightest subhalo galaxy.
                As the brightest cluster galaxy often lies at the center of the galaxy cluster in terms of both position and velocity, we check if a similar connection applies to subhalos.
                We define the brightest subhalo galaxy as the cluster member galaxy with the brightest $r$-band magnitude within $r_{t}$ of the subhalo center.
                Because the selection of brightest galaxy is sensitive to outliers and the dominant galaxy is expected to be close to the subhalo center, we use a stricter condition of $r_{t}$ instead of $2r_{t}$ used for GMM.
                The $p$-values are $p_{\mathrm{KS}}=18.7\%$ and $p_{\mathrm{AD}}=18.9\%$.
                Again, we cannot reject the null hypothesis that the cluster galaxies and subhalos are drawn from the same distributions.

                \begin{figure*}
                    \centerline{\includegraphics[width=1.0\linewidth]{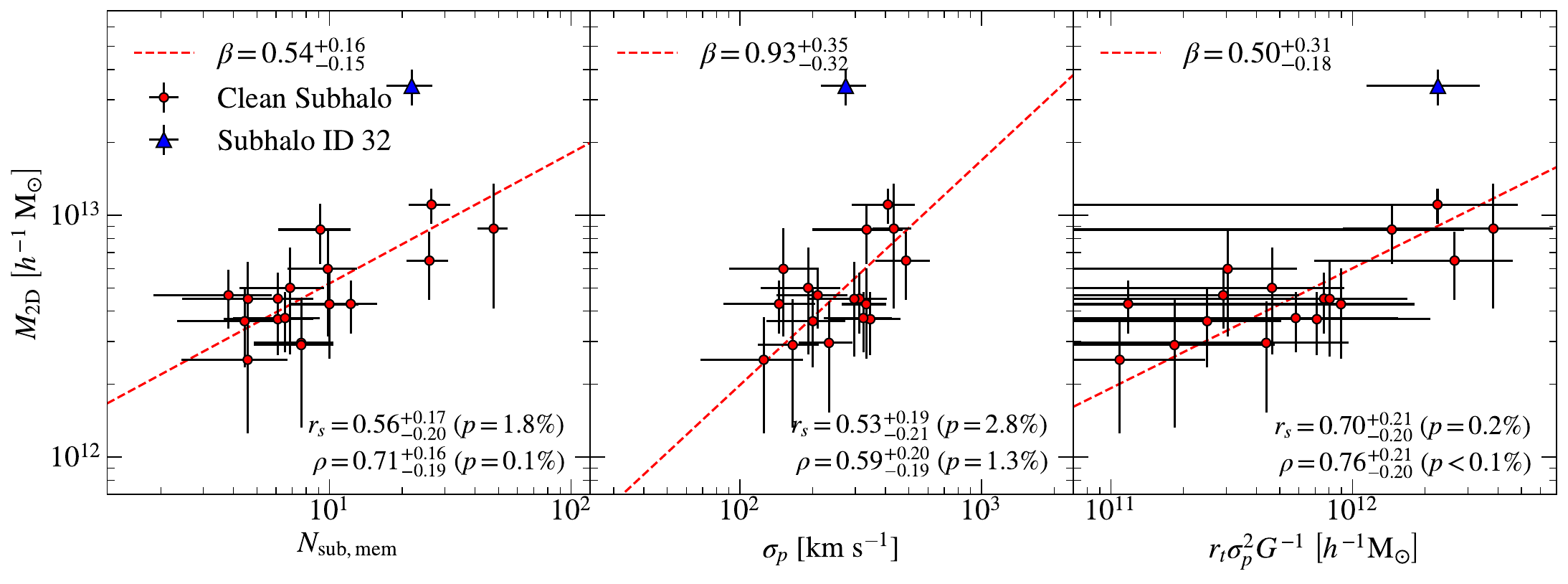}}
                    \caption{
                       Mass scaling relation of weak-lensing subhalos.
                       Left: Relation between weak-lensing mass $M_{\mathrm{2D}}$ and number of subhalo members $N_{\mathrm{mem}}$.
                       Center: Relation between $M_{\mathrm{2D}}$ and the line-of-sight velocity dispersion of galaxies within the subhalo $\sigma_{p}$.
                       Right: $M_{\mathrm{2D}}$ versus dynamical mass $r_{t} \sigma_{p}^{2}G^{-1}$.
                       The red circles represent the ``clean'' subhalos flagged for the GMM.
                       We also show Subhalo ID 32 for reference.
                       The red dashed lines show the best-fitting power laws only for the ``clean'' subhalos.               
                    }
                    \label{fig:mass_scale}
                \end{figure*}
    
            \subsubsection{Mass Scaling Relations}
                
                We investigate the scaling relations between the weak-lensing mass of subhalos $M_{\mathrm{2D}}$ and the physical quantities calculated from the galaxy redshift data using GMM.
                For virialized systems, one would expect that the mass of the system is proportional to the size of the system times the velocity dispersion squared.
                Motivated from the virial theorem, we use $r_{t}\sigma_{p}^{2}G^{-1}$ as a proxy of dynamical mass, where $G$ is the gravitational constant used for conversion into units of mass.
                In Figure \ref{fig:mass_scale}, we plot the relations between $M_{\mathrm{2D}}$ and $N_{\mathrm{sub,mem}}$ (left), $\sigma_{p}$ (middle), and $r_{t} \sigma_{p}^{2}G^{-1}$ (right).
                The red circles represent the ``clean'' subhalos.
                We also show the values for Subhalo ID 32 as the blue triangles.
                The mass of Subhalo ID 32 is much greater than other subhalos of similar physical quantities, which is likely a sign of contamination in $M_{\mathrm{2D}}$ due to the background cluster as discussed in Section \ref{sec:res:bkg}.
                We use the Poisson error as the error bars of $N_{\mathrm{sub,mem}}$, and the errors propagated from $\sigma_{p}$ and $r_{t}$ for the error bars of $r_{t}\sigma_{p}^{2}G^{-1}$.

                We fit the data to a power-law relation using only the ``clean'' subhalos.
                Specifically, we fit the data to a function
                \begin{equation}    \label{eq:linear}
                    \ln \frac{Y}{Y_\mathrm{piv}} = \alpha + \beta \ln \frac{X}{X_{\mathrm{piv}}}
                \end{equation}
                where $\ln$ is the natural logarithm.
                $X$ and $Y$ are either $M_{\mathrm{2D}}$, $N_{\mathrm{sub,mem}}$, $\sigma_{p}$, or $r_{t}\sigma_{p}^{2}G^{-1}$.
                We use the pivot values ($X_{\mathrm{piv}}$ and $Y_{\mathrm{piv}}$) $10^{12}h^{-1}\mathrm{M_{\Sun}}$ for $M_{\mathrm{2D}}$, $10$ for $N_{\mathrm{sub,mem}}$, $100\ \mathrm{km \ s^{-1}}$ for $\sigma_{p}$, and $10^{12}h^{-1}\mathrm{M_{\Sun}}$ for $r_{t}\sigma_{p}^{2}G^{-1}$.
                We employ the hierarchical Bayesian modeling for the linear regression used in Section \ref{sec:res:bkg}.
                The selection function is corrected for the case where there is an observational threshold in the detection in $Y$.
                In our case, the selection of samples is mostly dominated by the detection limit of the weak-lensing subhalos.
                As such, we focus on the case where $Y=M_{\mathrm{2D}}$, but present the results using other quantities as $Y$ for reference.

                \begin{deluxetable}{ccrcc}
                    \label{tab:scale}
                    \caption{Scaling relations of subhalos. Parameters are defined in Equations \ref{eq:linear}. The values for $\sigma_{\ln Y}$ are the 68\% upper limits.}
                    \tablehead{
                        \colhead{$X$} & \colhead{$Y$} & \colhead{$\alpha$} & \colhead{$\beta$} & \colhead{$\sigma_{\ln Y}$}
                    }
                    \startdata
                        $N_{\mathrm{sub,mem}}$      & $M_{\mathrm{2D}}           $ & $ 1.65^{+0.10}_{-0.10}$ & $0.54^{+0.16}_{-0.15}$ & $0.057$ \\ 
                        $M_{\mathrm{2D}}$           & $N_{\mathrm{sub,mem}}      $ & $-2.69^{+0.62}_{-0.74}$ & $1.63^{+0.41}_{-0.34}$ & $0.101$ \\ \hline
                        $\sigma_{p}$                & $M_{\mathrm{2D}}           $ & $ 0.68^{+0.36}_{-0.41}$ & $0.93^{+0.35}_{-0.32}$ & $0.080$ \\ 
                        $M_{\mathrm{2D}}$           & $\sigma_{p}                $ & $-0.05^{+0.44}_{-0.51}$ & $0.67^{+0.30}_{-0.26}$ & $0.060$ \\ \hline
                        $r_{t}\sigma_{p}^{2}G^{-1}$ & $M_{\mathrm{2D}}           $ & $ 1.80^{+0.21}_{-0.17}$ & $0.50^{+0.31}_{-0.18}$ & $0.066$ \\
                        $M_{\mathrm{2D}}$           & $r_{t}\sigma_{p}^{2}G^{-1} $ & $-2.72^{+1.47}_{-1.65}$ & $1.47^{+0.96}_{-0.84}$ & $0.106$ \\ 
                    \enddata
                \end{deluxetable}
                
                We summarize the results in Table \ref{tab:scale}, including $\alpha$, $\beta$, and the intrinsic scatter along $\ln Y$, $\sigma_{\ln Y}$.
                The errors are 68\% confidence intervals for $\alpha$ and $\beta$.
                The intrinsic scatter $\sigma_{\ln Y}$ is not constrained with our data, and thus we present the 68\% upper limit of $\sigma_{\ln Y}$.
                We note that one needs to take the inverse of the slope of $X=N_{\mathrm{sub,mem}}$, $Y=M_{\mathrm{2D}}$ to compare with the case of $X=M_{\mathrm{2D}}$, $Y=N_{\mathrm{sub,mem}}$, for example.
                In Figure \ref{fig:mass_scale}, the best-fitting power laws for $Y=M_{\mathrm{2D}}$ are represented in the red dashed lines.

                We also note that the measurement error in $r_{t}$ dominates the relation between $M_{\mathrm{2D}}$ and $r_{t}\sigma_{p}^{2}G^{-1}$.
                The conventional error propagation formula would not hold for estimating the error in $\ln (r_{t}\sigma_{p}^{2}G^{-1})$, as the formula requires the errors to be much smaller than the measured values.
                Thus, the confidence interval for $M_{\mathrm{2D}}$--$r_{t}\sigma_{p}^{2}G^{-1}$ should not be taken at face value.
                The relation would be better constrained when a more precise measurement of $r_{t}$ is given.

                The slope for $r_{t}\sigma_{p}^{2}G^{-1}$ is an expected byproduct of the power-law relation of $M_{\mathrm{2D}}$--$\sigma_{p}$ and $M_{\mathrm{2D}}$--$r_{t}$.
                Since $M_{\mathrm{2D}} \propto r_{t}$ and $M_{\mathrm{2D}} \propto \sigma_{p}^{0.93}$, we would expect $M_{\mathrm{2D}} \propto (r_{t}\sigma_{p}^{2})^{\sim 3}$, which is similar with our results within the uncertainty.
            
                We test the correlation between $X$ ($N_{\mathrm{sub,mem}}$, $\sigma_{p}$, and $r_{t}\sigma_{p}^{2}G^{-1}$) and $Y$ ($M_{\mathrm{2D}}$).
                We calculate the Spearman's rank correlation coefficient $r_{s}$ and Pearson's correlation coefficient $\rho$ and the corresponding $p$-values between $\ln X$ and $\ln Y$.
                The coefficients ($p$-values) are
                $r_{s}=0.56^{+0.17}_{-0.20}$ ($p=1.8\%$) and $\rho=0.71^{+0.16}_{-0.19}$ ($p=0.1\%$) for $M_{\mathrm{2D}}$--$N_{\mathrm{sub,mem}}$,
                $r_{s}=0.53^{+0.19}_{-0.21}$ ($p=2.8\%$) and $\rho=0.59^{+0.20}_{-0.19}$ ($p=1.3\%$) for $M_{\mathrm{2D}}$--$\sigma_{p}$, and
                $r_{s}=0.70^{+0.21}_{-0.20}$ ($p=0.2\%$) and $\rho=0.76^{+0.21}_{-0.20}$ ($p=0.04\%$) for $M_{\mathrm{2D}}$--$r_{t}\sigma_{p}^{2}G^{-1}$.
                The uncertainties are 68\% intervals, which we estimate by randomly sampling data points from lognormal distributions with the scales set to the measurement errors.
                These values are annotated in Figure \ref{fig:mass_scale}.
    
    \section{Discussion} \label{sec:discuss}
            
        \subsection{A Comparison with Subhalos in Galaxy Clusters from Cosmological Simulations}
            To better understand the results in this study, we anlayze the data from numerical simulations.
            Here, we would like to identify a Coma-like cluster in simulations and to check whether we could reproduce the results in the previous sections.
            In this way, we could examine the possible effects of different observational systematics, such as the misidentification of subhalo member galaxies due to the projection effect, while having access to the true three-dimensional (3D) information.

            \begin{figure*}[htb]
                \fig{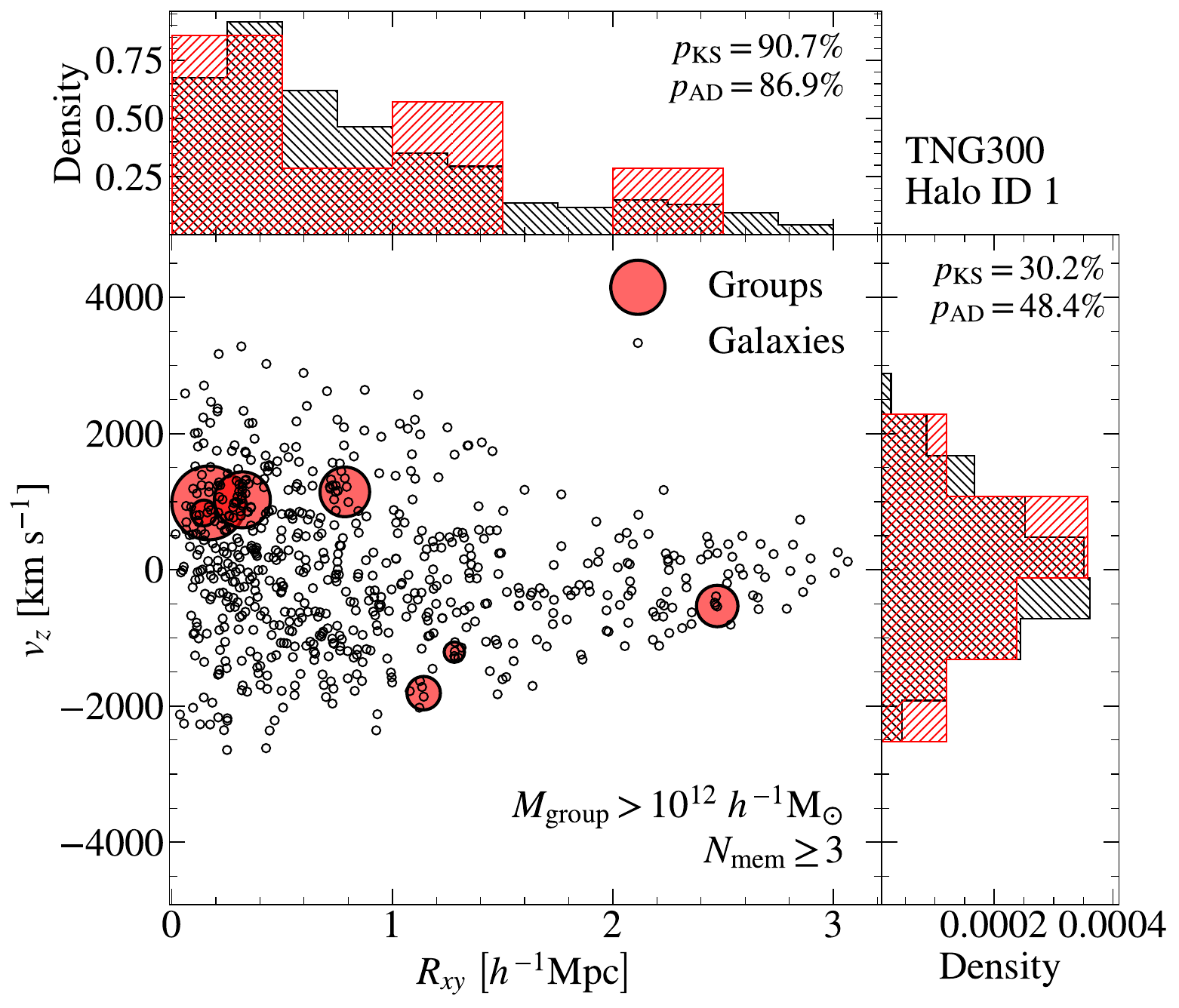}{1\columnwidth}{(a)}
                \fig{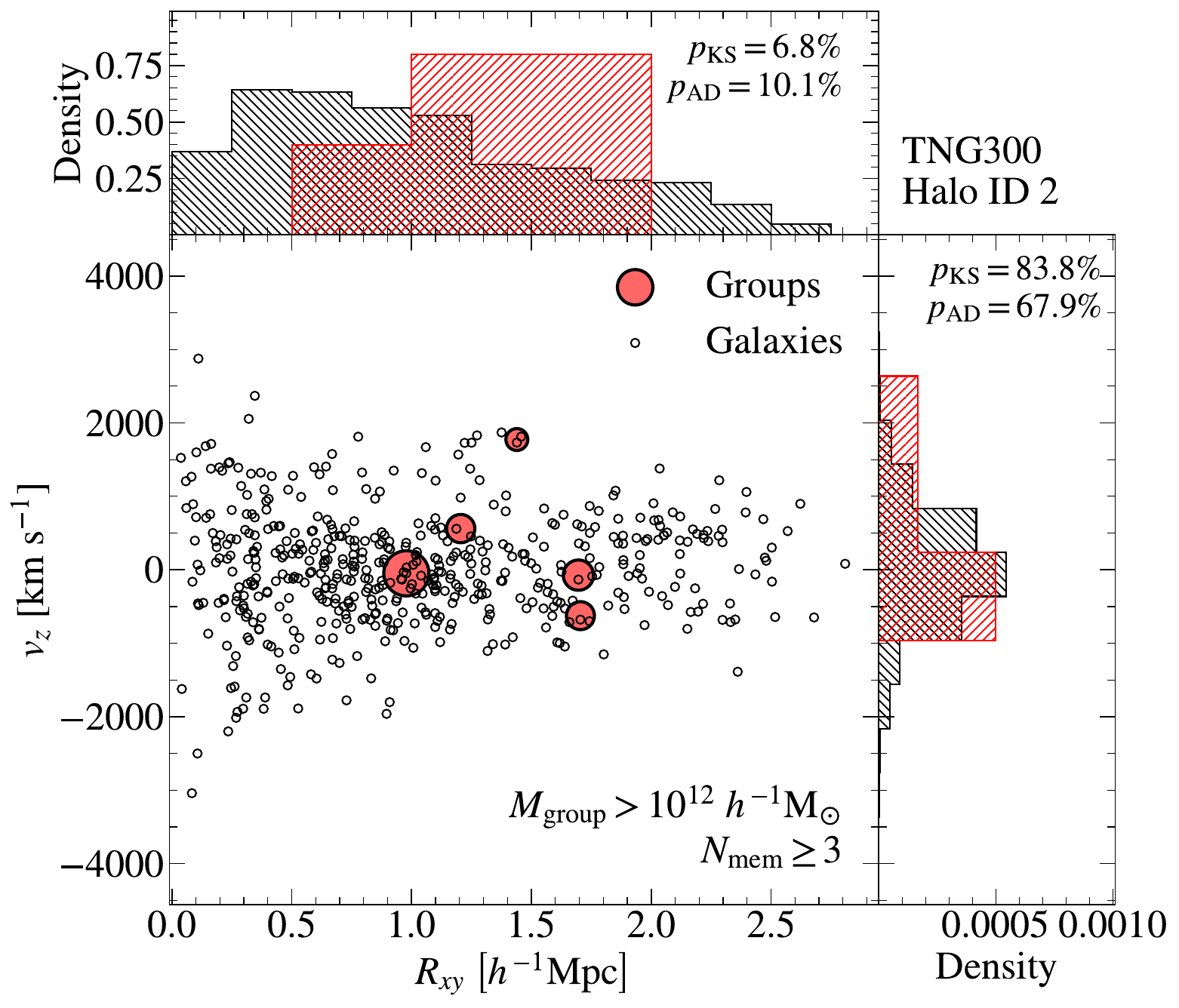}{1\columnwidth}{(b)}
                \caption{
                    Redshift space distribution of galaxies and subhalos within the clusters HaloID 1 (a) and Halo ID 2 (b) from the TNG300 simulation.
                    This figure is analogous to Figure \ref{fig:subhalo_caustic}.
                    Bottom left: Black open circles and red filled circles represent galaxies and groups with three or more members.
                    Top left and bottom right: Histograms of data points corresponding to the same colors in the bottom left panel.}
                \label{fig:tng_phase}
            \end{figure*}

            \subsubsection{A Coma-like Galaxy Cluster from IllustrisTNG}

                We use the results of the IllustrisTNG (TNG) project \citep{Marinacci2018_TNG, Naiman2018_TNG, Nelson2018_TNG, Pillepich2018_TNG300, Springel2018_TNG}.
                TNG is a series of cosmological magnetohydrodynamical simulation runs, with box sizes of 50 Mpc, 100 Mpc, and 300 Mpc.
                The results of each run are often referred to as TNG50, TNG100, and TNG300, respectively.
                The cosmological parameters assumed in the simulations are taken from  \citet{Planck2016_2015_CosmoParam}: $\Omega_m=0.3089$, $\Omega_b=0.0486$, $\Omega_{\Lambda}=0.6911$, $H_{0}=67.74\ \mathrm{km\ s^{-1}\ Mpc^{-1}}$, $\sigma_{8}=0.8159$, $n_{s}=0.9667$.
    
                It is important to clarify the terminology and notions used in the TNG simulations.
                The TNG data includes the position, velocity, and mass information of the particles (dark matter, star, and gas), as well as a catalog of halos found with the friends-of-friends (FoF) algorithm \citep{Huchra1982}.
                TNG also provides a catalog of subhalos within the FoF halos found with the \texttt{SUBFIND} algorithm \citep{Springel2001_subfind}.
                The mass scalses of FoF halos and \texttt{SUBFIND} subhalos are typically cluster- and galaxy-scale structures, respectively.
                Therefore, we interchangeably use the term galaxy cluster with FoF halo and galaxy with \texttt{SUBFIND} subhalos.

                We use snapshot number 99 of the TNG300, which corresponds to the redshift $z=0$.
                Among the four clusters with masses comparable to the Coma cluster (i.e., $M_{200} > 6\times10^{14} h^{-1} \mathrm{M_{\odot}}$) in TNG300, we use Halo IDs 1 and 2 as an example study.
                The other two clusters (Halo IDs 0 and 3) were excluded due to their merging activity.
                Therefore, the result from this analysis may not reflect the general properties of the clusters in TNG300.
                A thorough analysis with a larger sample of clusters needs to be done as a separate work (e.g., using the TNG-Cluster simulation of \citealp{Nelson202_TNG_Cluster}).

                The massive \texttt{SUBFIND} subhalos in the two TNG300 galaxy clusters are comparable to the weak-lensing subhalos in the Coma cluster.
                For example, there are 17 \texttt{SUBFIND} subhalos in Halo ID 1 with total mass greater than $10^{12} h^{-1} \mathrm{M_{\odot}}$ and 13 in Halo ID 2.
                Thus, we use the massive \texttt{SUBFIND} subhalos as the basis for the comparison with the observation results of the Coma cluster.
                To differentiate the terms referring to individual galaxies (\texttt{SUBFIND} subhalos) and substructures with similar mass scales to the weak-lensing subhalos, we refer the latter as groups.
                We calculate the mass of a group $M_{\mathrm{group}}$ as the sum of the mass of dark matter and stellar particles within $3r_{1/2}$ from the subhalo centers, where $r_{1/2}$ is the half-mass radius (i.e., the radius in which half of the total mass is enclosed).
                We assign galaxies within $3r_{1/2}$ to the groups and select galaxies brighter than $-15.8$ mag in $r$-band absolute magnitude.
                This absolute magnitude limit corresponds to apparent magnitude of 20.2 mag at redshift $z=0.230$.
                The host galaxy of the group is also included in the assigned galaxies.
                We count the number of group member galaxies brighter than $-15.8$ mag as $N_{\mathrm{mem}}$.

                \begin{figure*}
                    \centerline{\includegraphics[width=0.8\linewidth]{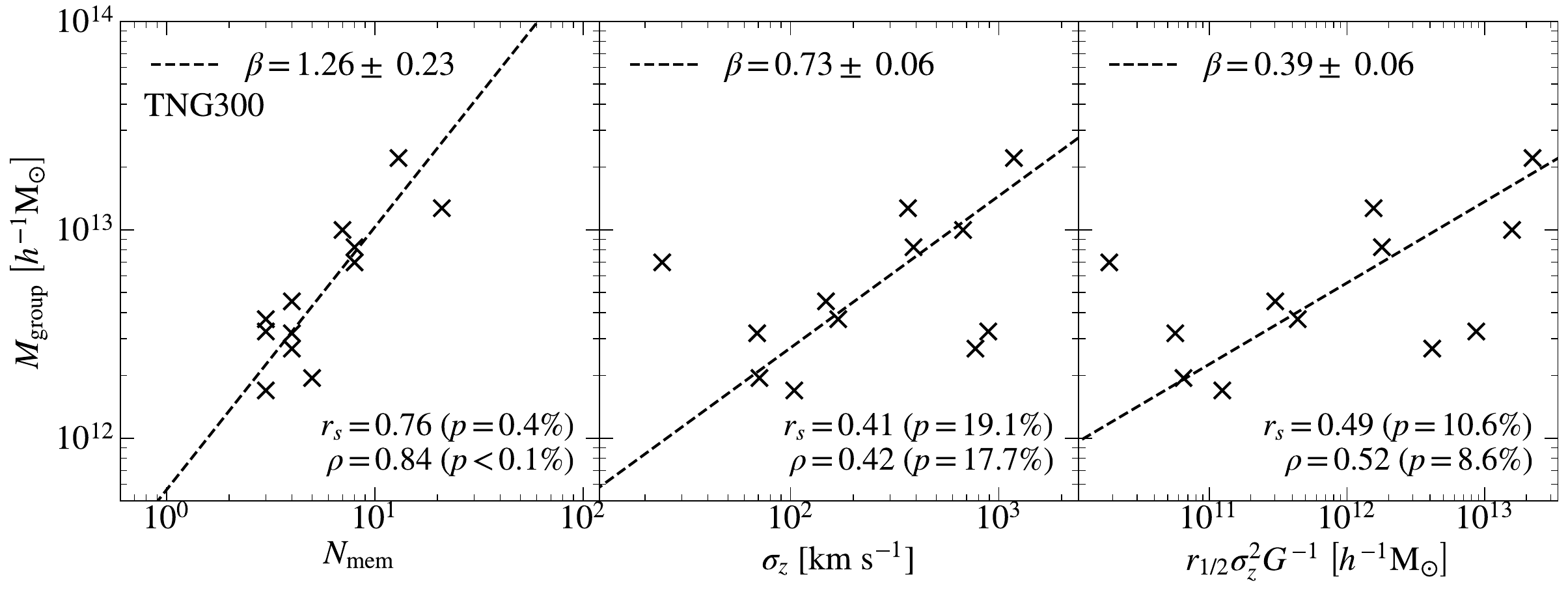}}
                    \caption{
                       Mass scaling relations of subhalos in two galaxy clusters in the TNG300 simulation.
                       The black crosses represent galaxy groups with three or more members and $M_{\mathrm{group}}>10^{12}\ h^{-1}\mathrm{M_{\odot}}$.
                       The black dashed lines show the best-fit power law.
                    }
                    \label{fig:tng_scale}
                \end{figure*}

            \subsubsection{Comparison of Galaxies and Groups in the Simulation}
                In Figure \ref{fig:tng_phase} (a) and (b), we compare the redshift space distribution of cluster member galaxies and groups in the Halo IDs 1 and 2 of the TNG300, respectively.
                The figure is analogous to Figure \ref{fig:subhalo_caustic}.
                In the bottom left panel, the black open circles and red circles represent cluster member galaxies and groups where the radii of the red circles are proportional to $M_{\mathrm{group}}$.
                We use the $xy$ plane of the simulation box as the sky plane and the $z$ axis as the line of sight.
                That is, the clustercentric distance is the projected distance in the $xy$ plane ($R_{xy}$) and the line-of-sight velocity is the velocity along the $z$ axis ($v_{z}$).
                Galaxies with $r$-band absolute magnitude brighter than $-15.8$ mag and groups with $M_{\mathrm{group}} > 10^{12} h^{-1} \ \mathrm{M_{\Sun}}$ and $N_{\mathrm{mem}} \geq 3$ are used.
                The line-of-sight velocity $v_{z}$ of a group is calculated as the average $z$-axis velocity of the group members.
                On the top left and bottom right panels of Figure \ref{fig:tng_phase}, the distribution of cluster member galaxies and groups are shown in black and red histograms, respectively.

                To quantitatively compare the distributions of cluster member galaxies and groups along $R_{xy}$ and $v_{z}$, we conduct a KS test and a two-sample AD test along the two axes.
                The $p$-values for the $R_{xy}$ distribution of galaxies and groups within Halo ID 1 (ID 2) are $p_{\mathrm{KS}}=90.7\% \ (6.8\%)$ for the KS test and $p_{\mathrm{AD}}=86.9\% \ (10.1\%)$ for the AD test.
                The $p$-values for the $v_{z}$ distribution of galaxies and groups within Halo ID 1 (ID 2) are $p_{\mathrm{KS}}=30.2\% \ (83.8\%)$ for the KS test and $p_{\mathrm{AD}}=48.4\% \ (67.9\%)$ for the AD test.
                Thus, we cannot reject the null hypothesis that the cluster member galaxies and groups are drawn from the same distributions of $R_{xy}$ and $v_{z}$.
                This may be because there really is no difference, or because the sample size is insufficient.

            \subsubsection{Mass Scaling Relations of Groups in Simulations}
                
                In Figure \ref{fig:tng_scale}, we show the mass scaling relations similar to Figure \ref{fig:mass_scale}.
                We use groups with $N_{\mathrm{mem}} \geq 3$.
                In the left, middle, and right panels, we plot the mass scaling relations for $N_{\mathrm{mem}}$, the projected velocity dispersion along the $z$-axis of the simulation box $\sigma_{z}$, and $r_{1/2} \sigma_{z}^{2}G^{-1}$, respectively.
                The black crosses and dashed lines represent the groups and the best-fit power-law relations.
                The slopes of the power-law are $\beta=1.26 \pm 0.23$ for $M_{\mathrm{group}}$--$N_{\mathrm{mem}}$, $\beta=0.73 \pm 0.06$ for $M_{\mathrm{group}}$--$\sigma_{z}$, and $\beta=0.39 \pm 0.06$ for $M_{\mathrm{group}}$--${r_{1/2}} \sigma_{z}^{2}G^{-1}$ relation.
                We also compute the Spearman's rank correlation coefficient $r_{s}$ and Pearson's correlation coefficient $\rho$ and their corresponding $p$-values between the logarithms of the quantities.
                We find that $N_{\mathrm{mem}}$ shows a correlation with $M_{\mathrm{group}}$ with $p$-values $0.4\%$ and $0.06\%$ for the Spearman's rank correlation and Pearson's correlation, respectively.
                A statistically significant correlation cannot be inferred for $\sigma_{z}$ and $r_{1/2} \sigma_{z}^{2}G^{-1}$, though the power-law slopes indicate that they have a positive relation on average.
                
            \subsubsection{What Can We Learn from the Groups in Simulation?}

                Using the TNG300 simulation data, we find that the groups with $M_{\mathrm{group}} > 10^{12} h^{-1} \mathrm{M_{\Sun}}$ show no significant difference in the distribution within the phase-space diagram compared to the cluster member galaxies (Figure \ref{fig:tng_phase}).
                This is consistent with the observation results of the Coma cluster (Figure \ref{fig:subhalo_caustic}).
                The similarity between the observation results of the Coma cluster and the TNG300 data implies that the line-of-sight velocities of the Coma cluster subhalos measured with GMM reproduce well the line-of-sight velocities, at least in a statistical sense.

                We have also examined the mass scaling relations of galaxy groups within the TNG300 clusters.
                Although the slopes are different from those of the weak-lensing subhalos in the Coma cluster, we do see a power-law relation with a positive slope from both the TNG simulation and the observation of the Coma cluster.
                The difference in slope and the larger scatter in the TNG300 results may be due to systematics of both the simulation and observation.
                For example, the groups we identified in the TNG300 can be physically different from the subhalos in the Coma cluster found with weak-lensing because of the different selection methods.
                In addition, while we used the 2D sky positions and line-of-sight velocity for estimating the physical properties of the Coma cluster subhalos, in the TNG300 simulation data we used the true 3D positions of galaxies for group membership.
                It would require a separate study to take into these systematics for a quantitative comparison.
                
                It should be noted that the use of the TNG300 galaxy clusters for our discussion is not a statistically rigorous analysis, as we only used two clusters from TNG300.
                In addition, the identified groups in the TNG300 clusters does not strictly represent the weak-lensing subhalos in the Coma cluster, due to the different identification of such substructures.
                We would like to emphasize that here we use the simulated clusters as a simple example of how different scaling relations may appear for subhalos in the cluster environment.
                A thorough study can be done using a large set of massive clusters ($> 10^{14}h^{-1} \ \mathrm{M_{\Sun}}$) with diverse merger history within a cosmological simulation in the future.
                One such examples is the TNG-Cluster simulation \citep{Nelson202_TNG_Cluster}, which includes 92 clusters with masses greater than $10^{15}\ \mathrm{M_{\Sun}}$ at $z=0$.

        \subsection{Mass Estimate of Subhalo ID 32}
            In Section \ref{sec:res:bkg}, we have identified an additional background galaxy cluster in the line of sight of Subhalo ID 32.
            As such, it is possible that the weak-lensing mass of Subhalo ID 32 is affected by the background contamination.
            Assuming that Subhalo ID 32 follows the best-fit scaling relations, we could infer the expected mass of the subhalo.
            We estimate the mass of Subhalo ID 32 and its 68\% confidence interval based on the posterior distributions of the parameters $\alpha$ and $\beta$ and assuming that $N_{\mathrm{sub,mem}}$ and $\sigma_{p}$ follow lognormal distributions.
            The estimated mass is $M_{\mathrm{2D}} = 7.87^{+1.60}_{-1.20} \times 10^{12} h^{-1} \mathrm{M_{\Sun}}$ from the scaling relation of $N_{\mathrm{sub,mem}}$ and $M_{\mathrm{2D}}=5.03^{+1.00}_{-0.86} \times 10^{12} h^{-1} \mathrm{M_{\Sun}}$ from $\sigma_{p}$.
            Because the intrinsic scatters of the scaling relations are not well constrained, the mass estimates do not take into account the intrinsic scatters.
            These mass estimates of Subhalo ID 32 are factors of 4--7 lower than the weak-lensing mass.
            The higher lensing efficiencies of the background structures have likely caused such contamination in the lensing signal.

        \subsection{$M$--$\sigma$ Relation Across Different Mass Scales}
            
            The velocity dispersion has traditionally been used as a tracer of mass across a wide range of mass scales.
            The scaling relation between the line-of-sight velocity dispersion of cluster galaxies and the mass of the cluster has been actively studied using both numerical simulations and observation data.
            For example, \citet{Munari2013_sigma_M_relation} showed that, using numerical simulations, the dark matter particles trace the mass of the cluster with $M_{200} \propto \sigma^{3}$, while galaxies trace the mass with a slightly shallower slope.
            Observationally, the $M_{200}$--$\sigma_{p}$ relation for galaxy clusters has been examined using mass measured from the Sunyaev--Zel'dovich (SZ) effect (e.g., \citealp{Ruel2014_sigma_M_SPT-SZ, Rines2016}) and from X-ray observation \citep{Ruel2014_sigma_M_SPT-SZ}.
            The scaling relations from observed clusters generally are in good agreement with the expectation from the simulation results (see also \citealp{Sohn_2020_ApJ_891_129S, Sohn_2022_ApJ_931_31S}).
            
            The velocity dispersion of stars in elliptical galaxies has a correlation with the luminosity of the galaxy and in turn the stellar mass \citep{Faber_1976_ApJ_204_668F}.
            \citet{Zahid2016_ApJ_832_203Z} studied the relation between the stellar mass $M_{*}$ and stellar velocity dispersion of galaxies using the observational data and found that $\sigma_{p} \propto M_{*}^{0.3}$.
            They also showed that, using stellar-mass-to-halo-mass relation \citep{Behroozi_2013_ApJ_770_57B} to convert $M_{*}$ to $M_{200}$, the $M_{200}$--$\sigma_{p}$ relation is in agreement with the extrapolation from the $M_{200}$--$\sigma_{p}$ relation of galaxy clusters.
            The stellar velocity dispersion can also be used to constrain the truncation radius and total mass of galaxy-scale subhalos within clusters \citep{Monna_2015_MNRAS_447_1224M, Monna_2017_MNRAS_465_4589M}.

            Although the $M$--$\sigma_{p}$ relation of individual galaxies and galaxy clusters follow $M \propto \sigma_{p}^{3}$ expected from numerical simulations, the relation seems to be in debate for galaxy group scales.
            For example, while \citet{Zhang_2022_AA_663A_85Z, Zhang_2024_ApJ_960_71Z} found that the $M_{200}$--$\sigma_{p}$ relation follows the expected relation with a power-law slope of $\sim 3$, others have found a shallower slope of $\sim2$ \citep{Han_2015_MNRAS_446_1356H, Viola_2015_MNRAS_452_3529V} or $\sim 1.5$ \citep{Rana_2022_MNRAS_510_5408R}.
            The exact mass range examined varies by studies, but typically ranges from $M_{200} = 10^{12}$ to $M_{200} = 10^{14}h^{-1}\mathrm{M_{\Sun}}$.
            These studies measure the mass of known galaxy groups by stacking lensing signals.

            \begin{figure}
                \centerline{\includegraphics[width=0.9\linewidth]{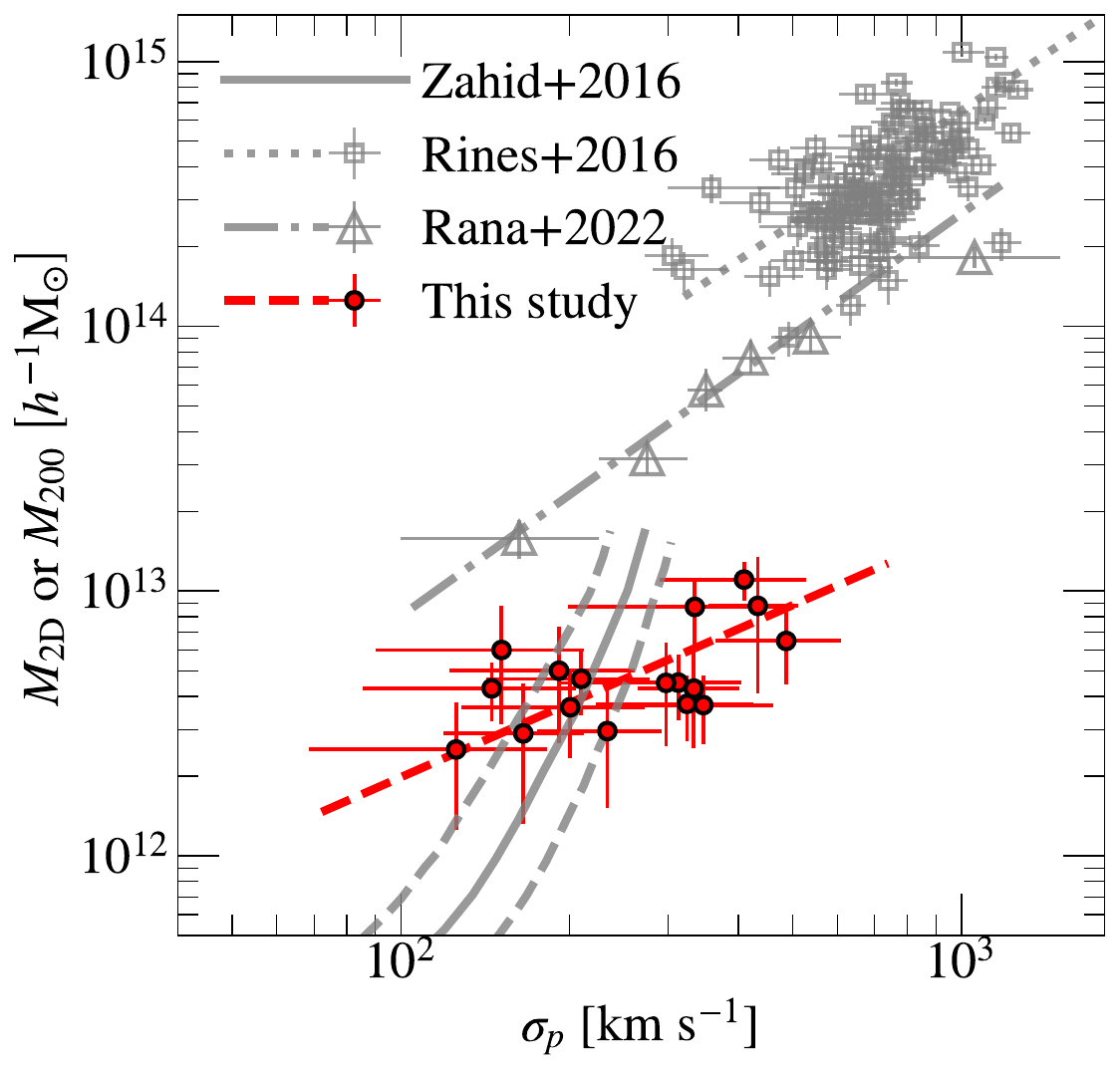}}
                \caption{
                   Comparison of $M$--$\sigma$ relation from the literature and this work.
                   The red circles and red dashed line represent the $M_{\mathrm{2D}}$--$\sigma_{p}$ relation from this work.
                   The dotted line and squares show the $M_{200}$--$\sigma_{p}$ relation of clusters from \citet{Rines2016}, and the dot-dashed line and triangles show the relation from \citet{Rana_2022_MNRAS_510_5408R}.
                   The gray solid line and dashed lines show the relation and its 68\% confidence interval for elliptical galaxies  \citep{Zahid2016_ApJ_832_203Z}.
                }
                \label{fig:M-sigma}
            \end{figure}

            In Figure \ref{fig:M-sigma}, we compare the $M$--$\sigma$ relation from the literature and this study, ranging from individual galaxies to clusters of galaxies.
            The $M_{\mathrm{2D}}$--$\sigma_{p}$ relation of subhalos in the Coma cluster from this study is shown in red circles with the best-fit power law in red dashed line.
            The $M_{200}$--$\sigma_{p}$ relation from \citet{Rines2016} and \citet{Rana_2022_MNRAS_510_5408R} are plotted as squares and triangles, with the best-fit power laws as dotted and dot-dashed lines, respectively.
            The $M_{200}$--$\sigma_{p}$ relation from \citet{Zahid2016_ApJ_832_203Z} is shown in the gray solid line, with the 68\% interval denoted as gray dashed lines.
            
            The $M$--$\sigma$ relations of different studies and different mass scales show diverse patterns.
            The power-law slope of the subhalos in the Coma cluster is the shallowest among the results shown in the figure.
            One possibility is the different definitions, and in turn the properties, of $M_{\mathrm{2D}}$ used in this study and $M_{200}$ in the other three studies.
            While $M_{200}$ is often used as the virial mass, $M_{\mathrm{2D}}$ is the total mass of the subhalo.
            In addition, the different environments in which the objects are situated could be the reason for the different slopes.
            The subhalos studied in this work lie in the cluster environment whose outer mass is stripped due to tidal interaction with the cluster gravitational potential \citep{Monna_2015_MNRAS_447_1224M, Monna_2017_MNRAS_465_4589M}.
            The average spherical density within $r_{t}$ of subhalos is much higher than 200 times the critical density at the Coma cluster's redshift, with the lowest average density being $\sim 10^4$ times the critical denstiy.
            In this regard, our findings are not in contradiction with previous studies.
            
            The $M$--$\sigma$ relation of weak-lensing subhalos presented this work is unique in the sense that it probes the group-scale mass of individual objects without binning or stacking, and that it is a direct study without relying on empirical relations.
            While we studied the $M$--$\sigma$ relation of weak-lensing subhalos in the Coma cluster, it is may also be possible to find galaxy groups within the Coma cluster first (e.g., \citealp{Adami_2005_A&A_443_17A, Jimenez_Teja_2024arXiv241215328J}) and then match with peaks in the weak-lensing map to study the group properties.
            Furthermore, the universality of the power-law relation found in this work remains to be answered.
            Identifying group-scale substructures within other clusters and examining the scaling relations would be required, which could be topics for future studies.

    \section{Conclusion} \label{sec:conclusion}
        We have combined the redshift data of our MMT/Hectospec observations of the Coma cluster with those from the literature to examine the physial properties of subhalos identified with weak-lensing analysis.
        The total number of galaxy redshifts is 12990 (\textcolor{black}{2183} from MMT/Hectospec observations and \textcolor{black}{10807} from the literature) in a 4.5 deg$^2$ region where the weak-lensing map exists.
        The magnitude limit where the differential spectroscopic completeness reaches 50\% is 20.2 mag in the $r$-band, which results in a high and spatially uniform completeness over the entire region of interest.
        We identify \textcolor{black}{1337} member galaxies of the Coma cluster using the caustic technique.
        We investigate weak-lensing group-scale subhalos found by \citet{Okabe2014} using the Coma member galaxies.
        We summarize our results as follows:
    
        \begin{enumerate}
            \item We measure the number of galaxies within the subhalos $N_\mathrm{sub,mem}$, the mean velocity of subhalo galaxies $v_{p}$, and the velocity dispersion of subhalo galaxies $\sigma_{p}$. We mitigate the effect of intracluster interlopers in the line of sight of the subhalos using GMM.
        
            \item The spatial and velocity distributions of subhalos are similar to the distributions of cluster member galaxies, as can be seen in Figure \ref{fig:subhalo_caustic}. This result is also expected from the TNG300 simulation data.
    
            \item The weak-lensing mass $M_{\mathrm{2D}}$ shows a power-law scaling relation with $N_{\mathrm{sub,mem}}$, $\sigma_{p}$, and $r_{t}\sigma_{p}^{2}G^{-1}$ (Figure \ref{fig:mass_scale}).
            The power-law slopes are $0.54^{+0.16}_{-0.15}$, $0.93^{+0.35}_{-0.32}$, and $0.50^{+0.31}_{-0.18}$, respectively.

            \item We identify an additional background object in the line of sight of Subhalo ID 32.
            Based on the scaling relation, we estimate the mass of Subhalo ID 32 to be $M_{\mathrm{2D}}=7.87^{+1.60}_{-1.20} \times 10^{12} h^{-1} \mathrm{M_{\Sun}}$ from $N_{\mathrm{sub,mem}}$ and $M_{\mathrm{2D}}=5.03^{+1.00}_{-0.86}\times 10^{12} h^{-1} \mathrm{M_{\Sun}}$ from $\sigma_{p}$.
    
            \item The slope of the $M$--$\sigma$ relation from this study is shallower than galaxy clusters ($\beta\sim3$), galaxy groups ($\beta \sim 1.5-3$), and individual galaxies ($\beta\sim3$), as shown in Figure \ref{fig:M-sigma}.
            This is likely due to not only the different definition of mass between $M_{\mathrm{2D}}$ and $M_{200}$, but also the environment in which the objects lie.
            
        \end{enumerate}

        To the best of our knowledge, this study is the first of its kind to study the nature of weak-lensing subhalos in clusters using galaxy redshift data.
        Our work underscores the importance of extensive redshift survey for studying structures within galaxy clusters.
        The comparison of weak-lensing data with the galaxy redshift data shown in this paper will be applicable to numerous other clusters with the data of current or upcoming projects, such as the full release of the DESI redshifts and the weak-lensing study to be led by the \textit{Euclid} mission \citep{Euclid2024_1}.

\section{Acknowledgments}
We thank the reviewer for constructive comments that helped us improve the manuscript.
We also thank M. James Jee and Kim HyeongHan for useful discussions.
H.S.H. acknowledges the support of the National Research Foundation of Korea (NRF) grant funded by the Korea government (MSIT), NRF-2021R1A2C1094577, Samsung Electronic Co., Ltd. (Project Number IO220811-01945-01), and Hyunsong Educational \& Cultural Foundation.
C.P. is supported by KIAS Individual Grants (PG016903) at Korea Institute for Advanced Study.
Funding for the Sloan Digital Sky Survey IV has been provided by the Alfred P. Sloan Foundation, the U.S. Department of Energy Office of Science and the Participating Institutions. SDSS-IV acknowledges support and resources from the Center for High-Performance Computing at the University of Utah. The SDSS web site is [www.sdss.org](http://www.sdss.org/).
    
SDSS-IV is managed by the Astrophysical Research Consortium for the Participating Institutions of the SDSS Collaboration including the Brazilian Participation Group, the Carnegie Institution for Science, Carnegie Mellon University, the Chilean Participation Group, the French Participation Group, Harvard-Smithsonian Center for Astrophysics, Instituto de Astrofísica de Canarias, The Johns Hopkins University, Kavli Institute for the Physics and Mathematics of the Universe (IPMU) / University of Tokyo, Lawrence Berkeley National Laboratory, Leibniz Institut für Astrophysik Potsdam (AIP), MaxPlanck-Institut für Astronomie (MPIA Heidelberg), Max-PlanckInstitut für Astrophysik (MPA Garching), Max-Planck-Institut für Extraterrestrische Physik (MPE), National Astronomical Observatories of China, New Mexico State University, New York University, University of Notre Dame, Observatário Nacional / MCTI, The Ohio State University, Pennsylvania State University, Shanghai Astronomical Observatory, United Kingdom Participation Group, Universidad Nacional Autónoma de México, University of Arizona, University of Colorado Boulder, University of Oxford, University of Portsmouth, University of Utah, University of Virginia, University of Washington, University of Wisconsin, Vanderbilt University and Yale University.

This research used data obtained with the Dark Energy Spectroscopic Instrument (DESI). DESI construction and operations is managed by the Lawrence Berkeley National Laboratory. This material is based upon work supported by the U.S. Department of Energy, Office of Science, Office of High-Energy Physics, under Contract No. DE–AC02–05CH11231, and by the National Energy Research Scientific Computing Center, a DOE Office of Science User Facility under the same contract. Additional support for DESI was provided by the U.S. National Science Foundation (NSF), Division of Astronomical Sciences under Contract No. AST-0950945 to the NSF’s National Optical-Infrared Astronomy Research Laboratory; the Science and Technology Facilities Council of the United Kingdom; the Gordon and Betty Moore Foundation; the Heising-Simons Foundation; the French Alternative Energies and Atomic Energy Commission (CEA); the National Council of Science and Technology of Mexico (CONACYT); the Ministry of Science and Innovation of Spain (MICINN), and by the DESI Member Institutions: www.desi.lbl.gov/collaborating-institutions. The DESI collaboration is honored to be permitted to conduct scientific research on Iolkam Du’ag (Kitt Peak), a mountain with particular significance to the Tohono O’odham Nation. Any opinions, findings, and conclusions or recommendations expressed in this material are those of the author(s) and do not necessarily reflect the views of the U.S. National Science Foundation, the U.S. Department of Energy, or any of the listed funding agencies.

\vspace{5mm}
\facility{MMT}

\software{
    Astropy \citep{astropy:2013, astropy:2018, astropy:2022},
    Matplotlib \citep{matplotlib},
    NumPy \citep{numpy},
    Scikit-learn \citep{scikit-learn},
    SciPy \citep{scipy}
}

\appendix
\section{Online-only Figure Sets}
\begin{figure}
    \centering
    \includegraphics[width=1\linewidth]{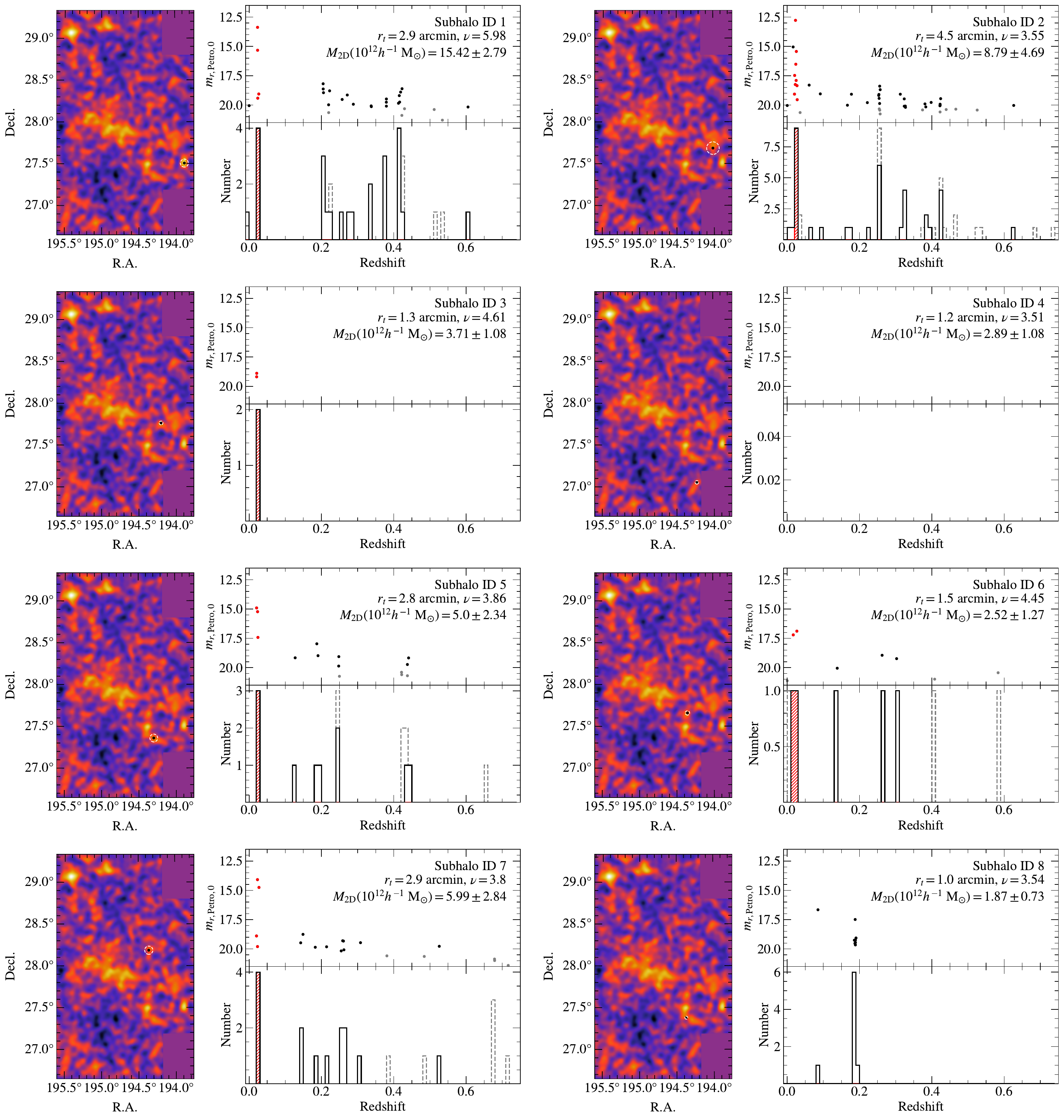}
    \caption{Same as Figure \ref{fig:subhalo-zhist}, for Subhalo IDs 1--8.}
    \label{fig:enter-label}
\end{figure}

\begin{figure}
    \centering
    \includegraphics[width=1\linewidth]{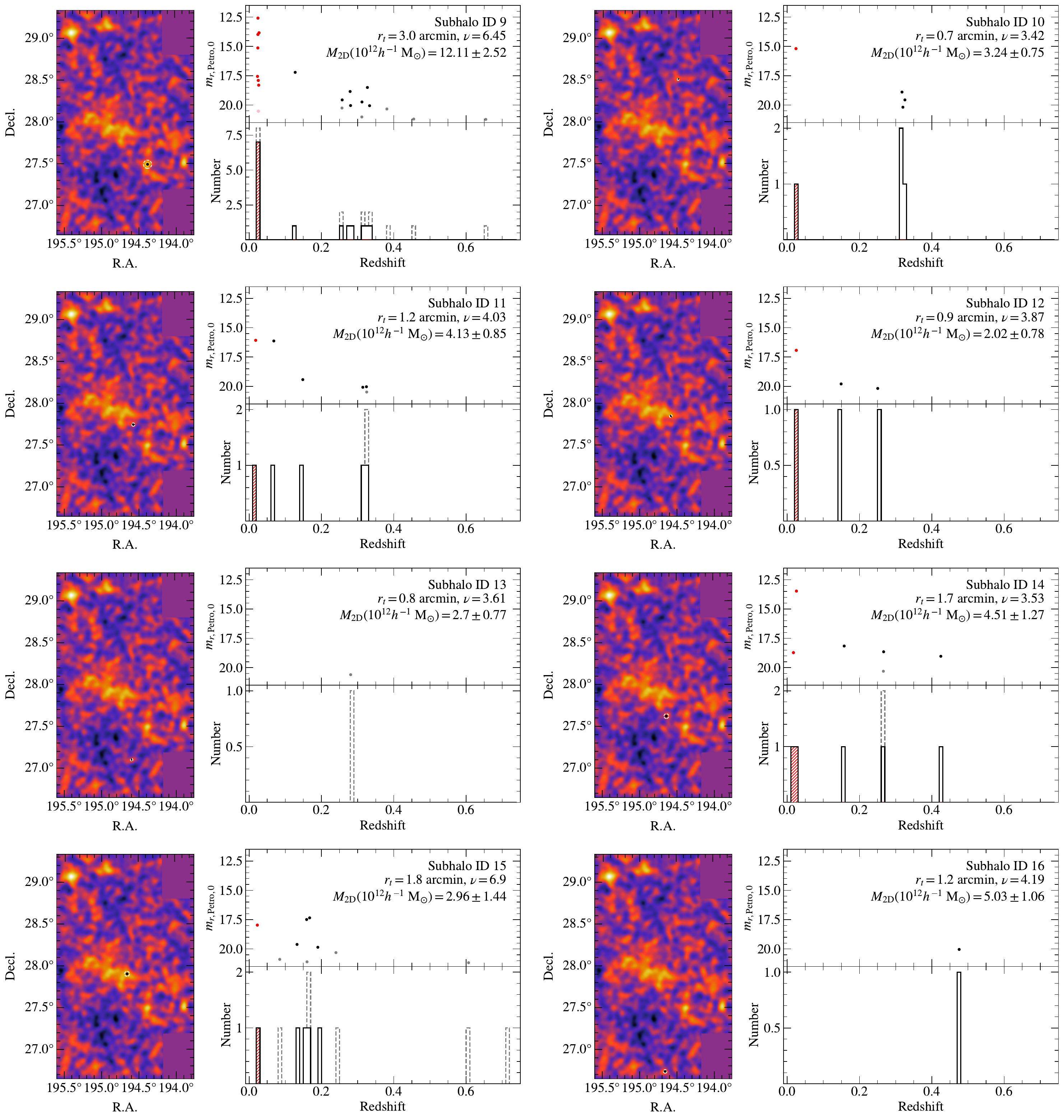}
    \caption{Same as Figure \ref{fig:subhalo-zhist}, for Subhalo IDs 9--16.}
    \label{fig:enter-label}
\end{figure}

\begin{figure}
    \centering
    \includegraphics[width=1\linewidth]{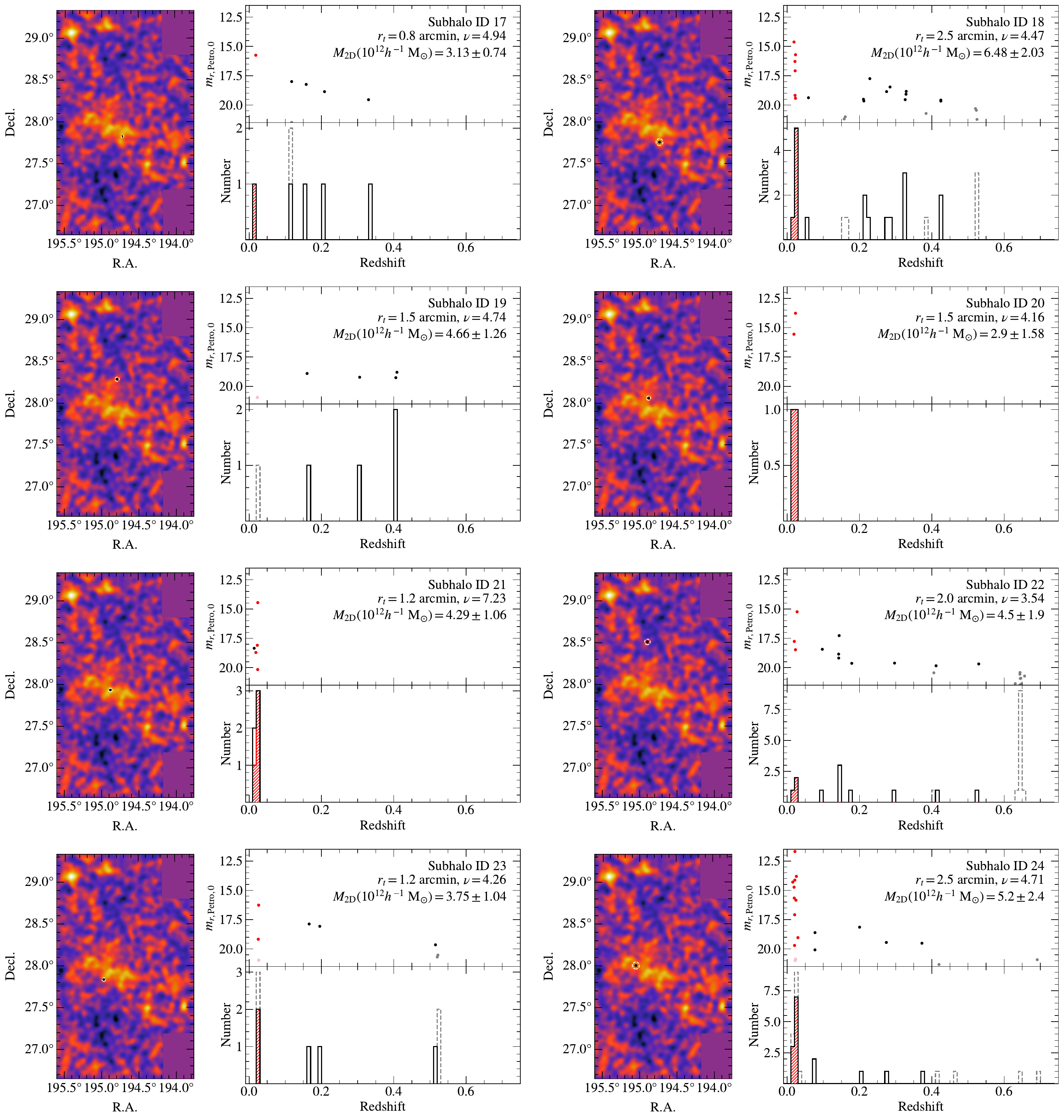}
    \caption{Same as Figure \ref{fig:subhalo-zhist}, for Subhalo IDs 17--24.}
    \label{fig:enter-label}
\end{figure}

\begin{figure}
    \centering
    \includegraphics[width=1\linewidth]{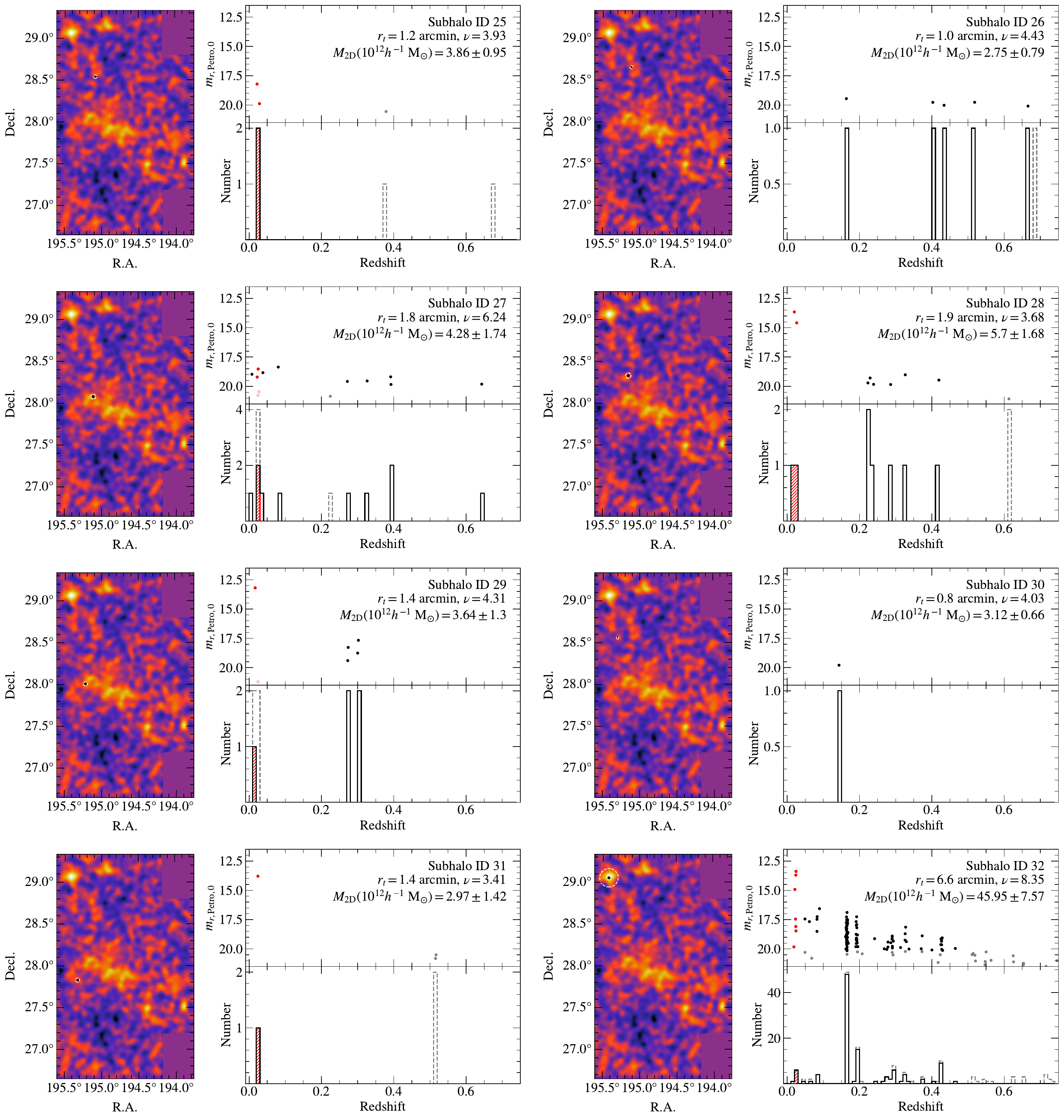}
    \caption{Same as Figure \ref{fig:subhalo-zhist}, for Subhalo IDs 25--32.}
    \label{fig:enter-label}
\end{figure}

\begin{figure}
    \centering
    \includegraphics[width=1\linewidth]{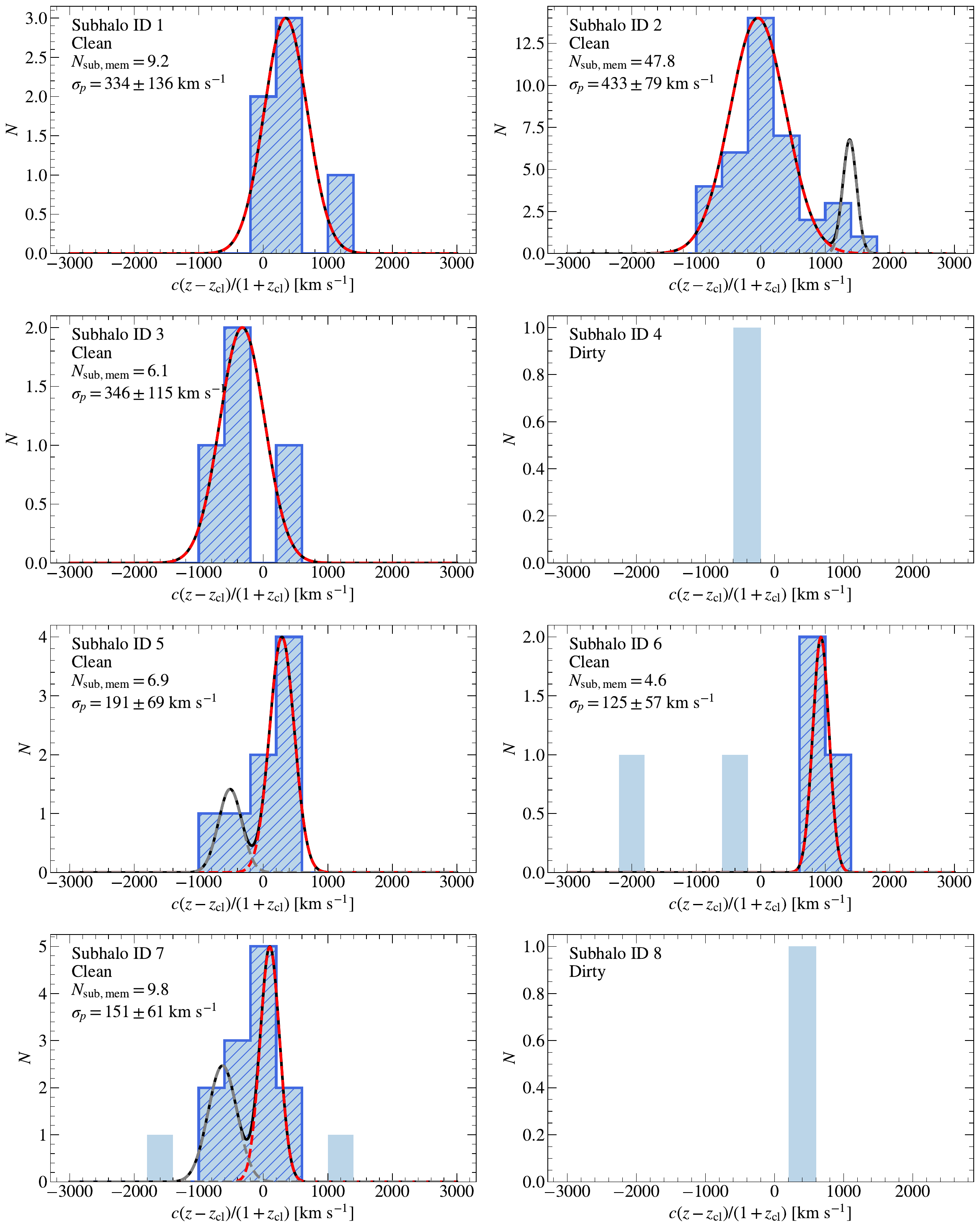}
    \caption{Same as Figure \ref{fig:gmm}, for Subhalo IDs 1--8.}
    \label{fig:enter-label}
\end{figure}

\begin{figure}
    \centering
    \includegraphics[width=1\linewidth]{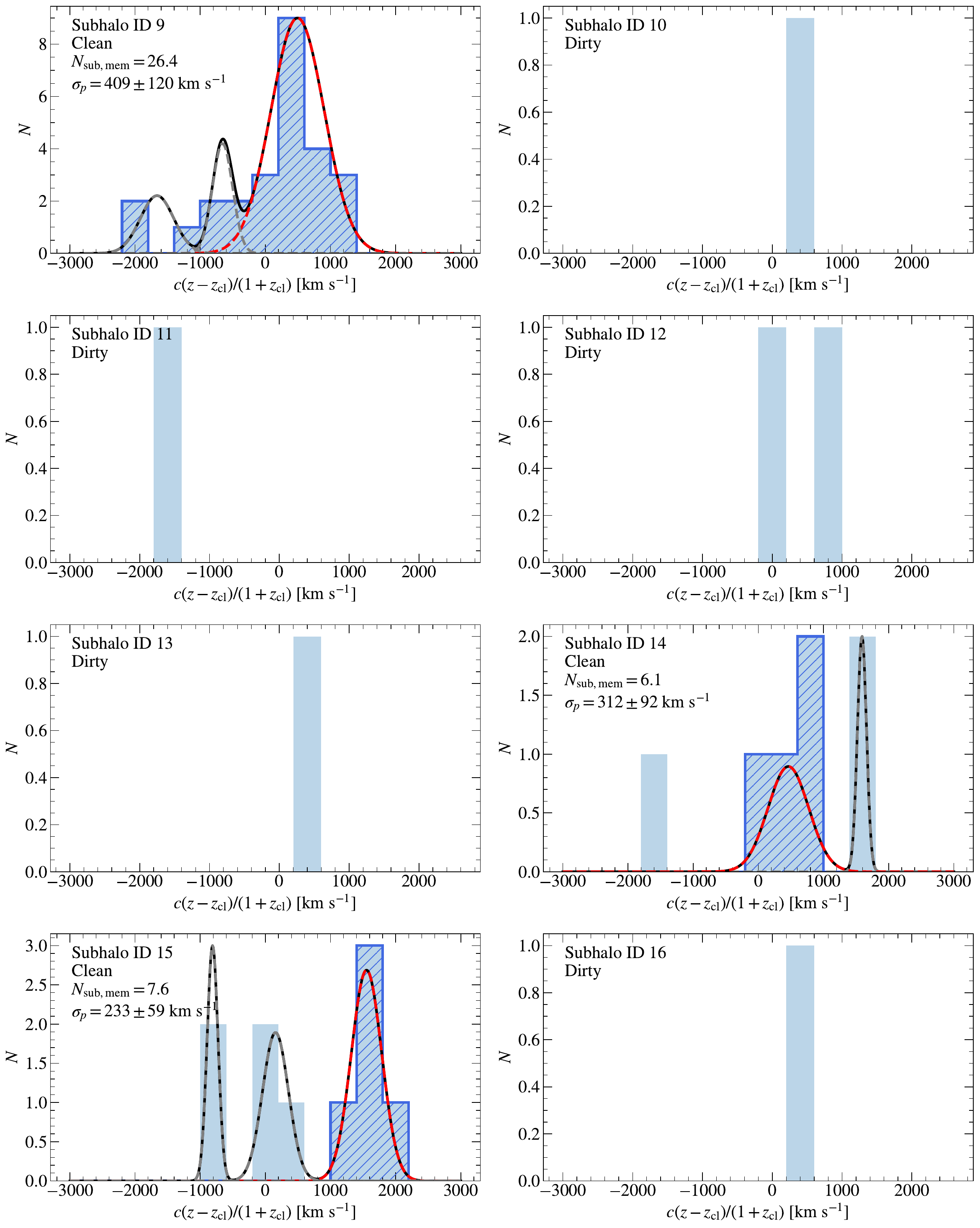}
    \caption{Same as Figure \ref{fig:gmm}, for Subhalo IDs 9--16.}
    \label{fig:enter-label}
\end{figure}

\begin{figure}
    \centering
    \includegraphics[width=1\linewidth]{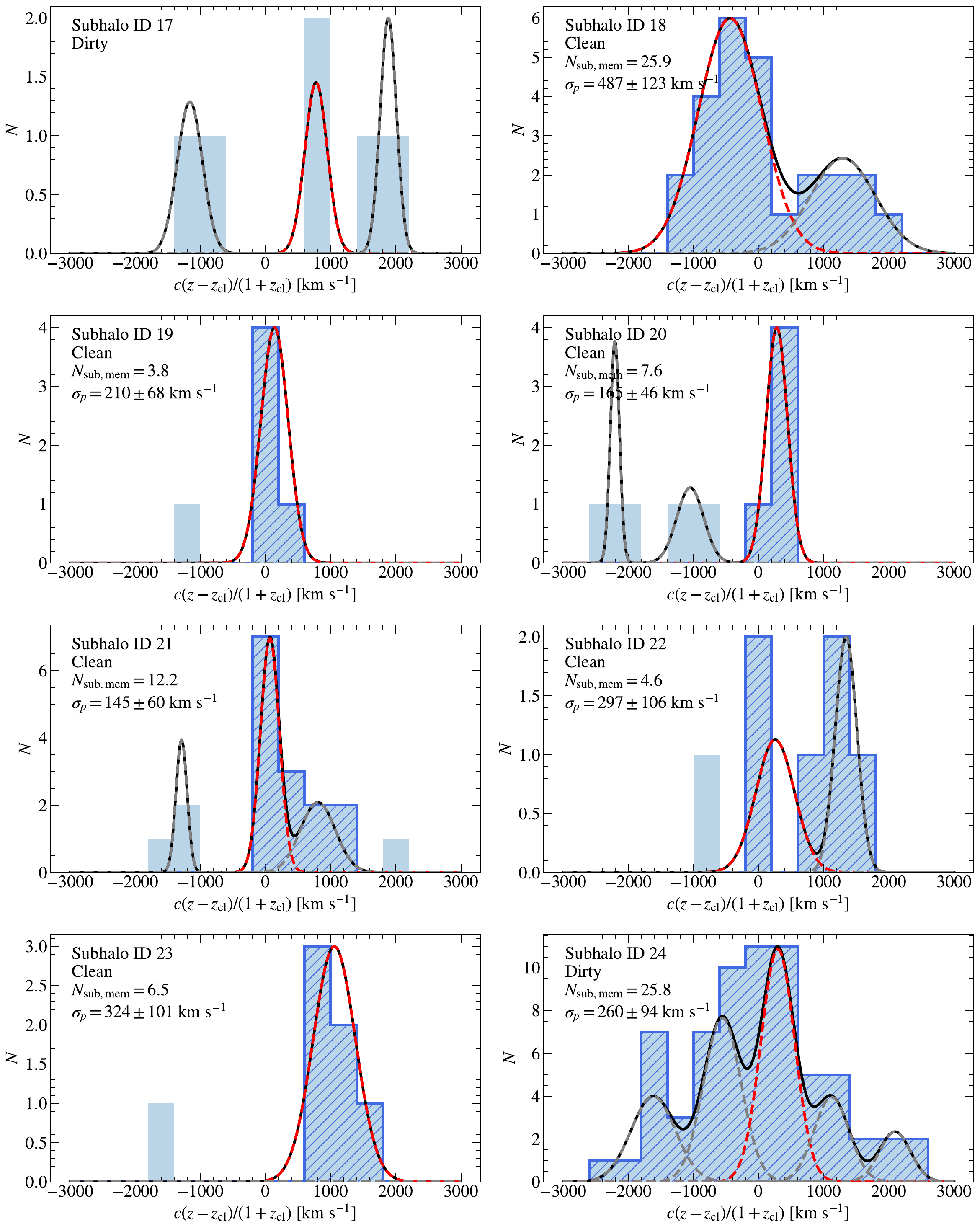}
    \caption{Same as Figure \ref{fig:gmm}, for Subhalo IDs 17--24.}
    \label{fig:enter-label}
\end{figure}

\begin{figure}
    \centering
    \includegraphics[width=1\linewidth]{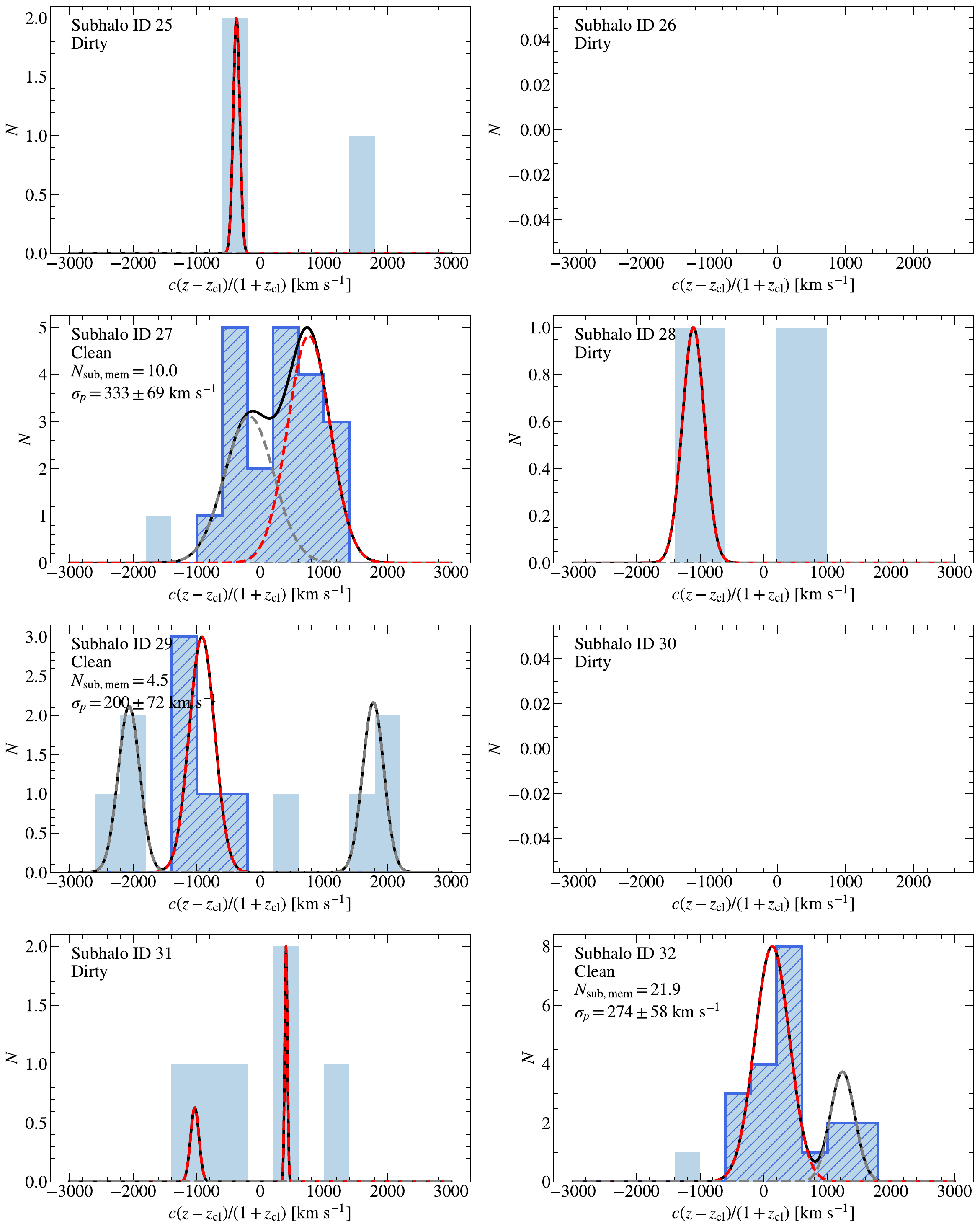}
    \caption{Same as Figure \ref{fig:gmm}, for Subhalo IDs 25--32.}
    \label{fig:enter-label}
\end{figure}

\bibliography{ref}{}
\bibliographystyle{aasjournal}

\end{document}